# Evolution of the Dust Trail of Comet 17P/Holmes

Maria Gritsevich[1,2,3,4], Markku Nissinen[2], Arto Oksanen[5], Jari Suomela[6], Elizabeth A. Silber[7,8]

1. Finnish Geospatial Research Institute (FGI), Vuorimiehentie 5, FI-02150 Espoo, Finland
2. Finnish Fireball Network, Ursa Astronomical Association, Kopernikuksentie 1, FI-00130 Helsinki, Finland
3. Department of Physics, University of Helsinki, Gustaf Hällströmin katu 2a, P.O. Box 64, FI-00014 Helsinki, Finland
4. Institute of Physics and Technology, Ural Federal University, street of Peace 19, 620002 Ekaterinburg
5. Hankasalmi observatory, Jyväskylän Sirius ry, Verkkoniementie 30, 40950 Muurame, Finland
6. Clayhole observatory, Tiriläntie 7, 05400 Tuusula, Finland
7. Department of Earth Sciences, Western University, London, ON, N6A 5B7, Canada
8. The Institute for Earth and Space Exploration, Western University, London, ON, N6A 3K7, Canada

The massive outburst of the comet 17P/Holmes in October 2007 is the largest known outburst by a comet thus far. We present a new comprehensive model describing the evolution of the dust trail produced in this phenomenon. The model comprises of multiparticle Monte Carlo approach including the solar radiation pressure effects, gravitational disturbance caused by Venus, Earth and Moon, Mars, Jupiter and Saturn, and gravitational interaction of the dust particles with the parent comet itself. Good accuracy of computations is achieved by its implementation in Orekit, which executes Dormad-Prince numerical integration methods with higher precision. We demonstrate performance of the model by simulating particle populations with sizes from 0.001 mm to 1 mm with corresponding spherically symmetric ejection speed distribution, and towards the Sun outburst modelling. The model is supplemented with and validated against the observations of the dust trail in common nodes for 0.5 and 1 revolutions. In all cases, the predicted trail position showed a good match to the observations. Additionally, the hourglass pattern of the trail was observed for the first time within this work. By using variations of the outburst model in our simulations, we determine that the assumption of the spherical symmetry of the ejected particles leads to the scenario compatible with the observed hourglass pattern. Using these data, we make predictions for the two-revolution dust trail behavior near the outburst point that should be detectable by using ground-based telescopes in 2022.

1. INTRODUCTION

During its recede from the Sun after the perihelion in 2007, comet 17P/Holmes underwent an enormous and sudden increase in brightness (Figure 1). This unique astronomical observation has been relatively well documented. The magnitude of the comet increased from a pre-outburst value of ~17 measured on 2007 October 23.1 (Casali et al. 2007) to 2.0 by 2007 October 25.1 (Sposetti et al. 2007; El-Houssieny, Nemiroff & Pickering 2010), which is equivalent to increase in brightness by a factor of one million (Hsieh et al. 2010). The outburst took place from October 23 to 24, 2007 (Montalto et al., 2008; Altenhoff et al., 2009; Lin et al., 2009; Sekanina et al., 2009; Reach et al., 2010; Hsieh et al., 2010; Kossacki et al., 2011; Ishiguro et al., 2013). It is the largest known outburst by a comet thus far recorded in the history of astronomical observations.

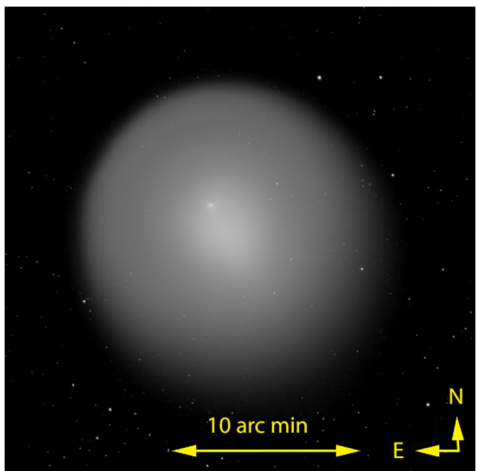

Figure 1. Comet 17P/Holmes observed at the Hankasalmi Observatory, Finland, on 2007 November 4 at 16:30:06 UT. The axes show the right ascension and declination. The compass shows north and east direction. Exposure time 60 seconds. Taken with SBIG ST-L-1001 CCD Camera.

A vast amount of dust particles and gas that were ejected from the comet's coma during the outburst spread into elliptic orbits around the Sun. This phenomenon and solar radiation pressure effect on the particles were investigated by Lyytinen et al. (2013). The evolving cloud of particles widened, apparently vanishing at first. However, Lyytinen et al. (2013) re-discovered this swarm of meteoroids, which converges again at the opposite side of the Sun around the mutual (southern) node of the orbits. In one revolution the particles re-converge again at the original outburst site (referred to in this study as the near-side common node or the northern node).

Due to the differences in the orbits, a passage of the particles through any of their nodes may take up to a year or even longer. Orbital period differences of the particles cause different node arriving times and differences in orbital elements cause the hourglass-shape. From purely gravitational modeling particle orbital periods can vary two years and the radiation pressure effect can lengthen orbital periods in theory to infinity for small particles (Lyytinen et al., 2013).

The increase in surface brightness of the trail of particles was expected to be sufficient to be observable in visible light. Therefore, Lyytinen et al. (2013) introduced the concept of direct observations of the ejected dust particles when they travel through the southern node situated farthest from the 2007 outburst point and later when they travel through the northern node situated at the 2007 outburst point itself.

In this study, we present results of the observations of the dust trail when it passes the both nodes. To the best of our knowledge, this is first direct observation of the hourglass pattern formed by the particles in the trail of a comet. Next, we present the new dust trail particle model, named the 'Dust Trail kit', implemented in Orekit, an open source space dynamics library (Maisonobe, Pommier and Parraud, 2010). The basis of the model is an improved model of propagation of the particles originally developed by Esko Lyytinen, which in this work's realization is coupled with a novel approach for simulating the outburst itself. The model comprises multiparticle Monte Carlo modeling, as well as gravitational disturbance caused by Venus, Earth and its Moon, Mars, Jupiter, Saturn, and gravitational interaction of the dust particles with the comet itself. The model is validated using the observational data obtained in the dedicated campaigns described in the following Section and in Supplementary material. Additionally, a search was conducted for the two-revolution trail between Fall 2020 and Spring 2021. The obtained results allow us to constrain the future behaviour of the dust trail of comet 17P/Holmes when it comes close again to the 2007 outburst point.

## 2. THEORETICAL CONSTRAINTS AND OBSERVATIONS

The near-side common node and the far-side common node differ greatly in particle dynamics. Significantly stronger convergence of the particles occurs in the near-side node. Convergence in the far side node is not well constrained in the direction of their distance from the Sun (Lyytinen et al., 2013). Therefore, observations of the dust particles in a vicinity of the near-side node are optimal. Moreover, after two revolutions, the dust particles remain near the 2007 outburst point (Lyytinen et al., 2013).

However, in practice, observations are possible for the particles converging in the far-side node when Earth is crossing the comet's orbital plane. This occurs two times a year. Comet 17P/Holmes was previously in perihelion on 19th February 2021. It takes half a year for the comet to travel to the 2007 outburst site. Prior to that, the orbital geometry is such that the comet and its dust trail are visible simultaneously in the same telescopic field of view.

The comet itself was aligned to the top of the dust trail also when observed from Earth in September 2014 (Ishiguro et al., 2016). This provided a rare opportunity to observe further spatial and temporal evolution of the dust trail. Below we provide a summary of the observations made within this project and further details are given in Supplementary Material.

The first observing campaign started in February of 2013, when the comet trail passed the southern node. We used telescope.net remote controlled telescopes in the Siding Spring Observatory, Australia. The observing campaign started immediately after the observatory was opened again after an unfortunate bush fire, which destroyed the visiting centre and caused problems to network communications in the observatory and in logistics. We decided to use mainly the Planewave 500mm Reflector with a CCD camera with Luminance filter for the observations. The image subtraction technique was used, where images taken in several nights are subtracted from each other.

For the observing campaign in the northern node that started in 2014 and continued into 2015, we used itelescope.net California, USA, and New Mexico, USA observatories with remote controlled telescopes. In California, we used the Planewave 610mm Reflector with CCD camera and New Mexico Planewave 413mm Reflector with CCD camera and Planewave 500mm Reflector with CCD camera, both with Luminance filter. We also used the Hankasalmi Observatory (Finland) remote controlled telescope with CCD camera and Luminance filter in 2014 and 2015. We used the image subtraction technique but after the trail brightened significantly, the image subtraction technique was no longer required.

Our observations revealed that the particle-cloud forms a cyclic "hourglass" pattern, which converges at specific points in space. These nodes are located where the particles' orbital planes cross each other. In the following section we describe how we also modeled this behavior.

3. DUST TRAIL MODELLING

The 2007 outburst process of the comet 17P/Holmes was studied in detail in the number of studies (Lin et al., 2009; Sekanina et al., 2009; Reach et al., 2010). Previous studies also detected and modeled the remnant dust cloud of comet 17P/Holmes (Ishiguro et al., 2016), as well as detected and modeled the dust cloud near the comet nucleus during the outburst (Hsieh et al., 2010).

It was proposed that the particle size can be modeled using the *β* parameter, which generates non-gravitational solar radiation pressure disturbance to the particle (Lyytinen et al., 2013). Ejection speed (*v*) was different for different sized particles (Reach et al., 2010). Smaller particles attained greater velocities than bigger particles. We have fixed 1 mm particles to have the solar radiation pressure effect *β* = 0.0002 after (Burns, Lamy and Soter, 1979; Landgraf et al. 2000). The particle-ejection speed relation and the corresponding *β* parameter relation are shown in Table 1. The list of all symbols used in our study is provided in Appendix.

We use 3 particle populations ranging from particulate to fine dust and referred to as follows: big (or larger) particles with 0.1 - 1 mm radius (named BPs), medium sized particles with 0.01 - 0.1 mm radius (MPs) and small particles with 0.001 - 0.01 mm radius (SPs). The distribution of the particles within each of the 3 groups is uniform. In the visible wavelengths range particulate dust is not observable at backscattering, at phase angles <30°, if it is composed of moderately or highly absorbing material (Zubko et al., 2013).

Table 1: Particle radius in mm, ejection velocity in m/s after (Reach et al., 2010) and the ratio of radiation pressure to gravity *β* after (Burns et al., 1979; Landgraf et al. 2000).

| r, mm | ejection velocity, m/s | β |
|---|---|---|
| 1 | 330 | 0.0002 |
| 0.1 | 515 | 0.0022 |
| 0.01 | 610 | 0.022 |
| 0.001 | 640 | 0.280 |

In this work we study the particles ejected during the 2007 outburst (Sekanina et al., 2009) from the coma of 17P/Holmes by building upon, developing, and applying the earlier version of the dust trail particle model (Lyytinen, 1999; Lyytinen and Van Flandern, 2000; Lyytinen, Nissinen and Van Flandern, 2001). The new model is implemented in Orekit, which allowed us to achieve the high accuracy of computations by executing Dormad-Prince numerical integration methods with higher precision. The following considerations were added to the 'Dust Trail kit' compared to the earlier versions of the model (Lyytinen, 1999; Lyytinen and Van Flandern, 2000; Lyytinen et al., 2001; Nissinen et al. 2021a):

1) Account for gravitational disturbances caused by Venus, Earth and Moon, Mars, Jupiter, and Saturn.
2) Inclusion of the particle's gravitational interaction with the parent comet.

3) Addition of the ejection speed section and 'particle feeding' as a one-particle-at-a-time algorithm using the Monte Carlo method.

The number of particles that can be considered in the model is arbitrary. Monte Carlo runs are accomplished with the Hipparchus add-in package to Orekit (Maisonobe et al., 2010).

We validated the model using the telescopic observations obtained from 2013 to 2015 (see details of the observations in Supplementary Material). The observations were done in both common nodes of particles' orbits (Lyytinen et al., 2013; 2014; 2015; Nissinen et al., 2021b), and are summarized in Table S1. These observations were compared to the modeled position, position angle, width of the trail and brightness fit of the observed trail as well as the modeled trail particle's integrated distribution.

Non-gravitational forces acting on the particles in a comet trail are well explained in (Vaubaillon, Colas, and Jorda, 2005). The solar radiation pressure is the result of the electromagnetic radiation emitted by the Sun exerted upon the particles. Other active forces are the Poynting and the (diurnal) Yarkovsky-Radzievskii effects produced by the anisotropy of the thermal radiation from the particles.

Following Lyytinen (1999), Lyytinen and Van Flandern (2000), and Lyytinen et al. (2001), the non-gravitational continuous acceleration parameter in the model is set as A2 = 0. The particle ejection speed distribution is assumed to be spherically symmetric to comet's coma. The ejection speed of the 0.001 mm particles in the model is 640 m/s, which is decreased with increasing the particle size (Reach et al., 2010), Table 1.

In addition, we also modelled particles having their ejection direction (only) towards the Sun with the same physical model, the same ejection velocity distribution, and the corresponding $β$ parameter relation.

The outburst time is not exactly known as different studies provide different, sometimes inconsistent estimates. A likely outburst date window of $t_0$= 2007 October 23.3 ± 0.3 was given by Hsieh et al. (2010) and 2007 October 24.5 in (Lin et al., 2009). The comet's starting point in our simulations was set as October 2007 23.5.

The simulations were performed with Orekit Open Source Library for Operational Flight Dynamics in the International Celestial Reference Frame (ICRF/J2000) standard celestial reference system. The position and velocity are calculated in standard m and m/s units. The simulations are executed in variable time steps. Spherically symmetric ejection velocity distribution is calculated using Sphere Point Picking method (Muller, 1959; Marsaglia, 1972). The used propagator is Runge-Kutta based Dormand-Prince integrator. The orbital elements of the comet for year 2007 are used from Giorgini (2021). Planetary and lunar locations are used as they are computed in Orekit, with the JPL Planetary and Lunar Ephemerides DE440 (Park et al., 2021).

For illustrating results of the model, the geocentric right ascension (RA) and declination (DEC) are calculated for observation time in the Earth mean equatorial coordinate system, and in the ICRF/J2000 reference frame and the coordinate system.

Based on the analysis of the expansion of the comet's coma, Hsieh et al. (2010) suggested that the outer coma was dominated by the material ejected in an instantaneous, explosive manner. In agreement with this, the 2007 outburst event is modeled as an impulse in our work. Reach et al. (2010) concluded that the outburst duration was less than 3 hours and had fast event rise time and slower decay time.

As shown in Table 1, the solar radiation pressure effect described by $\beta$ is effectively the particle size. The solar radiation pressure affected gravitational parameter of the Sun effective to the particle ($\mu'$) is used as an input in Orekit and is calculated using the gravitational parameter of the Sun ($\mu$) as described by (Burns et al., 1979):

$$\mu' = (1 - \beta)\mu \qquad (1)$$

The particles have a calculated location and velocity at the start of the modeling sequence. The particle ejection velocity distribution is spherically symmetric or towards the Sun and is scaled with corresponding $\beta$ value. The ejection velocity is added to the particle velocity at the start location:

$$\begin{bmatrix} v_x \\ v_y \\ v_z \end{bmatrix} = \begin{bmatrix} v_{0x} \\ v_{0y} \\ v_{0z} \end{bmatrix} + \begin{bmatrix} v_{ex} \\ v_{ey} \\ v_{ez} \end{bmatrix} \qquad (2)$$

When sampling the particles in the observation window and calculating RA and DEC in geocentric coordinate system, the Earth's location is subtracted:

$$\begin{bmatrix} x_{gw} \\ y_{gw} \\ z_{gw} \end{bmatrix} = \begin{bmatrix} x_p \\ y_p \\ z_p \end{bmatrix} - \begin{bmatrix} x_{earth} \\ y_{earth} \\ z_{earth} \end{bmatrix} \qquad (3)$$

For comparing the model with the observations, the declination $\alpha$ and right ascension $\theta$ are calculated using Orekit. The light-time correction, i.e. the displacement in the apparent position of an object from its geometric position caused by the object's motion during the time it took the light to reach an observer, was not taken into account in our calculations, although this may be a valid correction to consider (Dumoulin, 1994).

The coordinates of the comet and the Earth's position are taken from the JPL Horizons system (Giorgini, 2021). The trail spatial location is obtained from the dust trail model described in this Section.

4. RESULTS AND DISCUSSION

A typical simulation executed in this work is run with ~2000 particles in spherical outburst modelling and with ~800 particles in towards the Sun outburst modelling. When compared to observations, we select particles fitting for every case via random sampling using Hipparchus package and Orekit. Our visualization selection constraint is 40° variation in RA. Because the model gives out particle coordinates with one day resolution at JD zero decimal for clarity, the observed position of the trail is corrected by adjusting the time of the observation to match the timestamp used in the modelled coordinate list (Lyytinen et al., 2013, 2014).

The gravitational interaction of particles with the comet itself is included in the new model, although in our modelling results for this kind of outburst scenario its contribution was found to be negligible.

Below we discuss comet 17P/Holmes dust trail behavior starting from 2012.

4.1. Overall dust trail evolution from 2012 to 2021

At the beginning of 2012, the dust particles arrived near the southern node after orbiting 0.5 revolution past the 2007 outburst point. The medium sized particles arrived first, in a wide (few hundredths of AU) formation already in February 2021. In April 2012, the larger particles arrived. A few months later, in the summer or 2012, the small particles arrived, in a wide pattern distributed mainly at the orbit radii. The small particles compose the wide tail of the trail, dispersing wider even after the other particles had already left the southern node.

After completing one revolution past the outburst point, the dust particles arrived at the northern node at the beginning of 2014. The medium sized particles arrived first, followed by the larger particles immediately thereafter. The small particles arrived at the 2007 outburst point considerably later, towards the end of 2014.

In October 2014, the dust particles left the southern node. The long tail of the trail with small particles was moving outside of the orbit radii. When small particles were leaving the vicinity of the southern node, they were ~0.3 AU farther in the Z heliocentric axis.

In October 2016, the majority of dust particles left the northern node 2007 outburst point, leaving behind only the small particles.

The larger and medium sized dust particles arrived again to the northern node two revolutions after the 2007 outburst, in June 2020. Physically, at that time, the spearhead of the tail was located approximately 0.1 AU further in the orbit radii. The small particles arrived later, in March 2021; when they arrived, the trail was spatially more near the 2007 outburst point compared to when the larger particles arrived. The density of small particles is low when they start populating the 2007 outburst point in March 2021.

4.2. Observations at the southern orbit node 2013 and 2014

We made three observing runs for the southern node. The dust had traveled to the southern node over a year before the observations started.

Our first observations of the dust were made at the Siding Spring Observatory in February 2013. The modeling results show that all particle sizes modeled were still present in the dust trail (Figures 2-5). The observed part of the dust trail was already situated towards the end part of the trail.

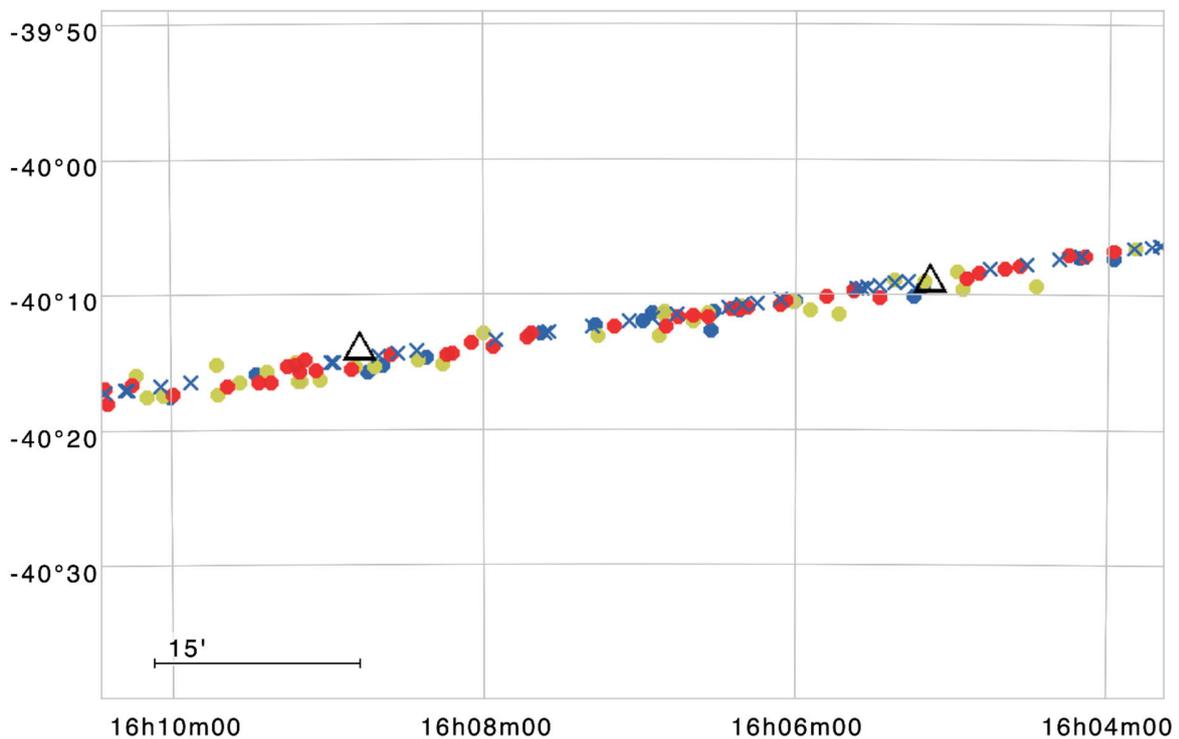

Figure 2. The 'Dust Trail kit' modeling for 2013 February 17 output is consistent with observations of (M2). X-axis shows RA and Y-axis DEC. The colour coding is used to illustrate different size particles. Blue: SPs. Yellow: MPs. Red: BPs. Black triangles: observed start and end positions of the trail. Particles ejected in the model towards the

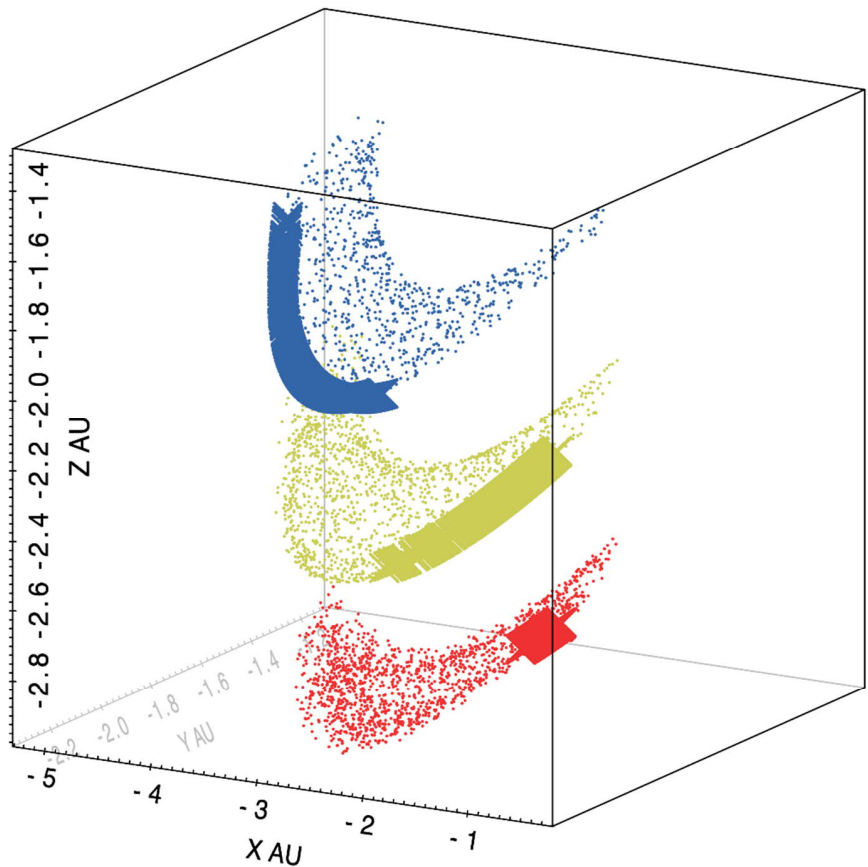

Figure 3. Modeling results for 2013 February 17 (at the time of observation M2). Particles are shown in ICRF coordinates XYZ. The colour coding used reflects the size of the modelled particles: Blue: SPs. Yellow: MPs. Red: BPs. In order to fully demonstrate all particle populations, we have applied offset Z = 0.5 to the blue particles and offset Z = -0.5 to the red particles. Particles ejected towards the Sun are marked with crosses.

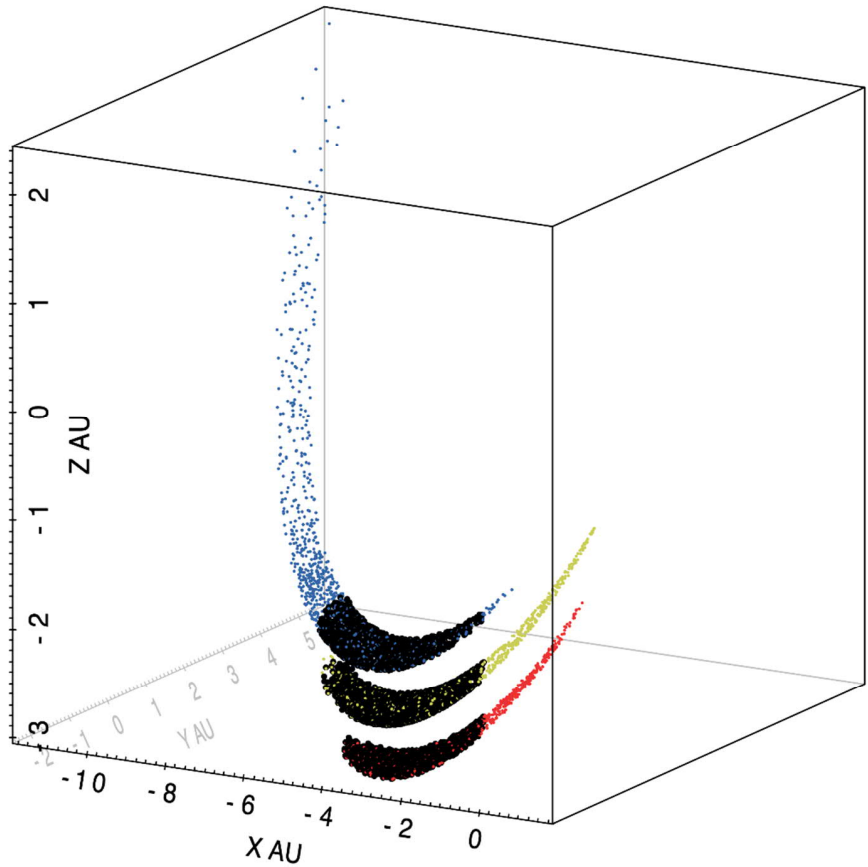

Figure 4. Modeling results for 2013 February 17 (at the time of observation M2). The particles are shown in ICRF coordinates XYZ. The colour code used reflects the size of the modelled particles: Blue: SPs. Yellow: MPs. Red: BPs. In order to fully demonstrate all particle populations, we have applied offset Z = 0.5 to the blue particles and offset Z = -0.5 to the red particles. Black circles: particles in the 40° RA window.

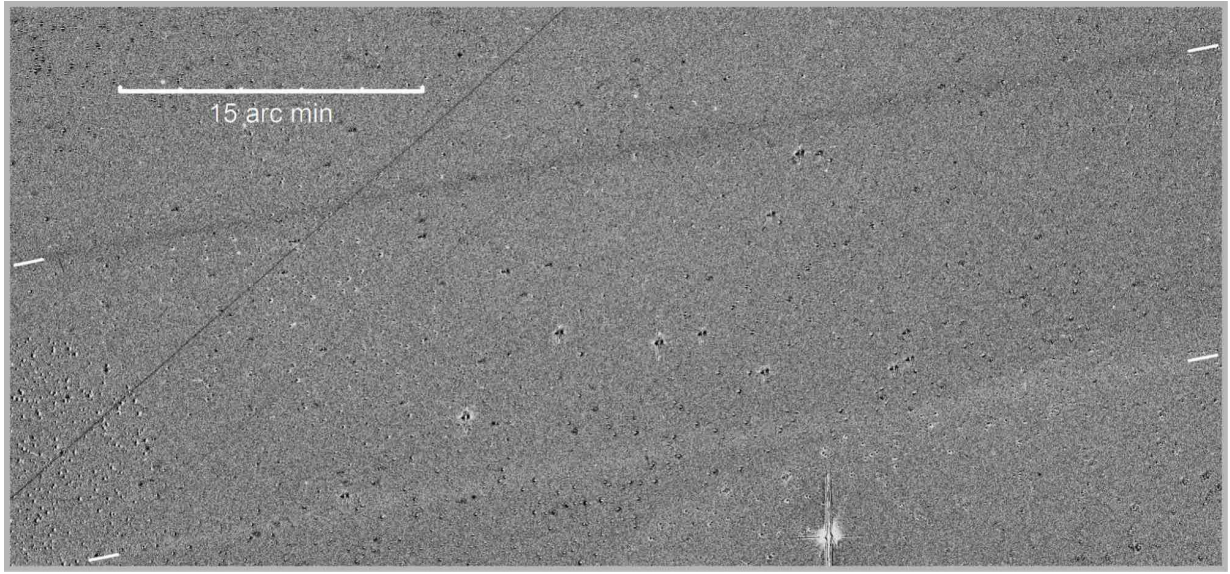

Figure 5. Observation made on 2013 February (M2). Darker trail is 17 February observation. Lighter trail is 19 February observation. Adopted from Lyytinen et al. 2013.

The second observation made in August 2013 showed a dust trail, which according to our model had still all sized particles present (Figures 6-7).

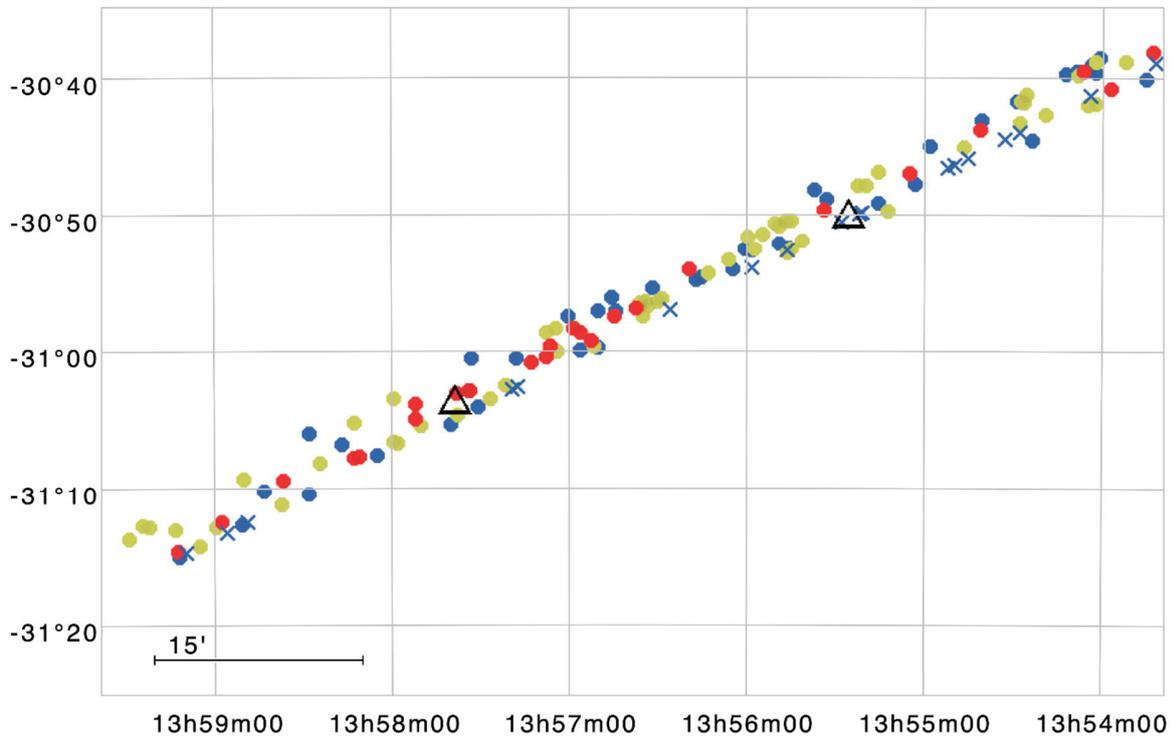

Figure 6. The 'Dust Trail kit' modeling for 2013 August 24 output is consistent with the observations of (M3). X-axis shows RA and Y-axis DEC. The colour coding is used to illustrate different size particles. Blue: SPs. Yellow: MPs. Red: BPs. Black triangles: the observed start and end positions of the trail. Particles ejected towards the Sun are marked with crosses.

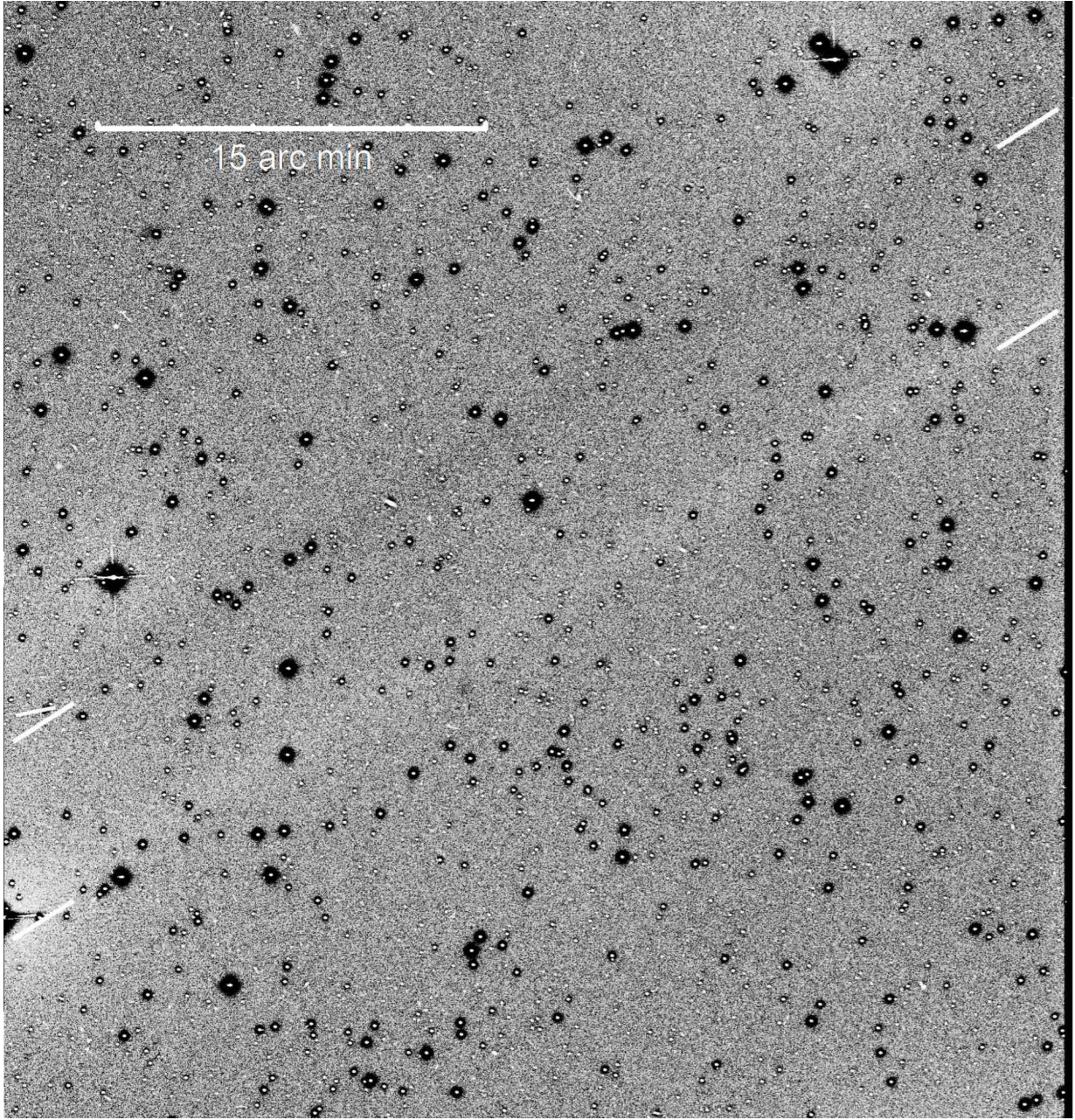

Figure 7. Observation of the 17P trail made in 2013 August 24 (M3). Image subtraction. The lighter trail is M3, see Supplementary material.

The third series of observations was performed in February 2014, and it showed only medium and small particles present in the dust trail (Figures 8-9).

Figure 8. Modelled results of observation 2014 February 11 (M4). X-axis shows RA and Y-axis DEC. Only small and medium sized particles are seen in the trail. Blue: SPs. Yellow: MPs. Black triangles: start and end observed positions of the trail. Our result indicates that it was possible to measure only small and medium sized particles in the southern node. Particles ejected towards the Sun are marked with crosses.

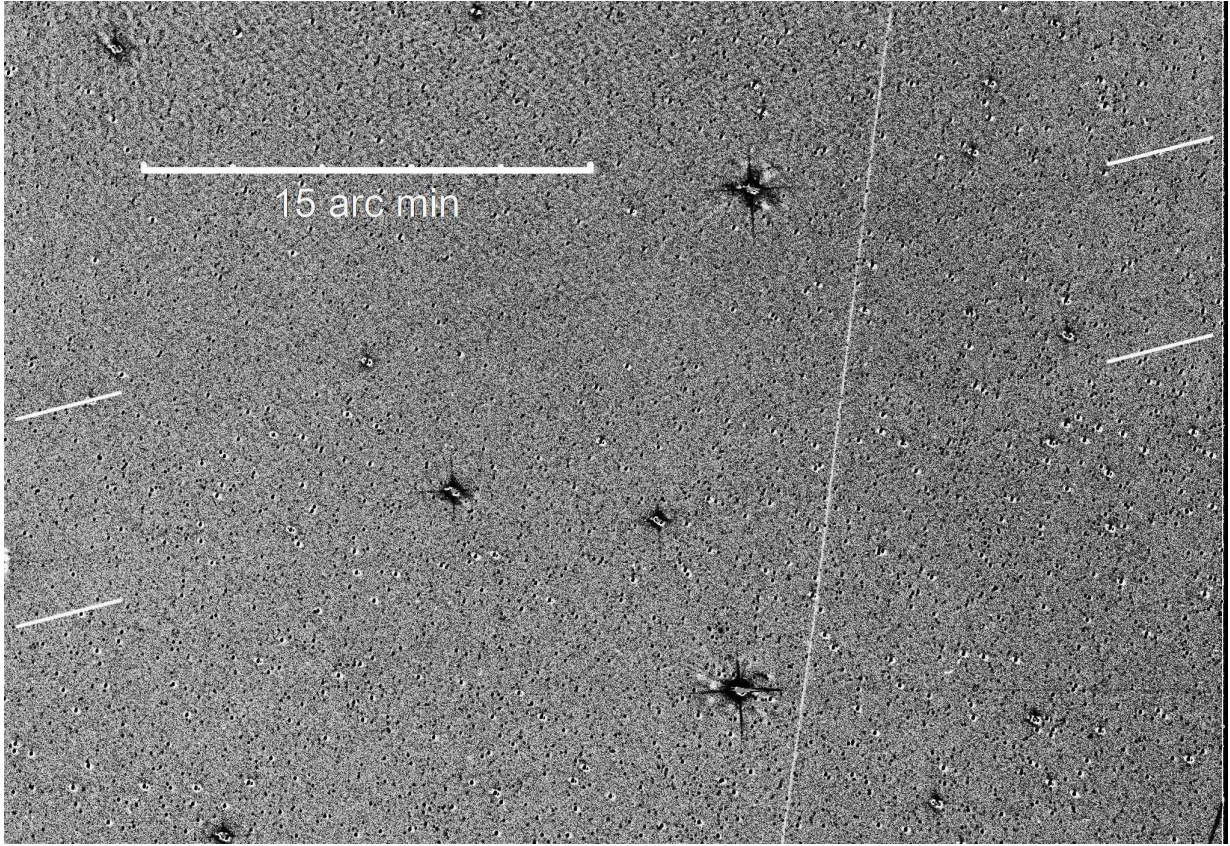

Figure 9. Observation made in 2014 February 11 (M4). Image subtraction. Upper trail is M4.

4.3. Observations near the 2007 outburst point at the northern orbit node 2014 and 2015

The dust had traveled to the 2007 outburst site for over six months before the first observation. Observations started in August 2014 from the Auberry Sierra Remote Observatory. All particle sizes were present in the dust trail (Figures 10-11).

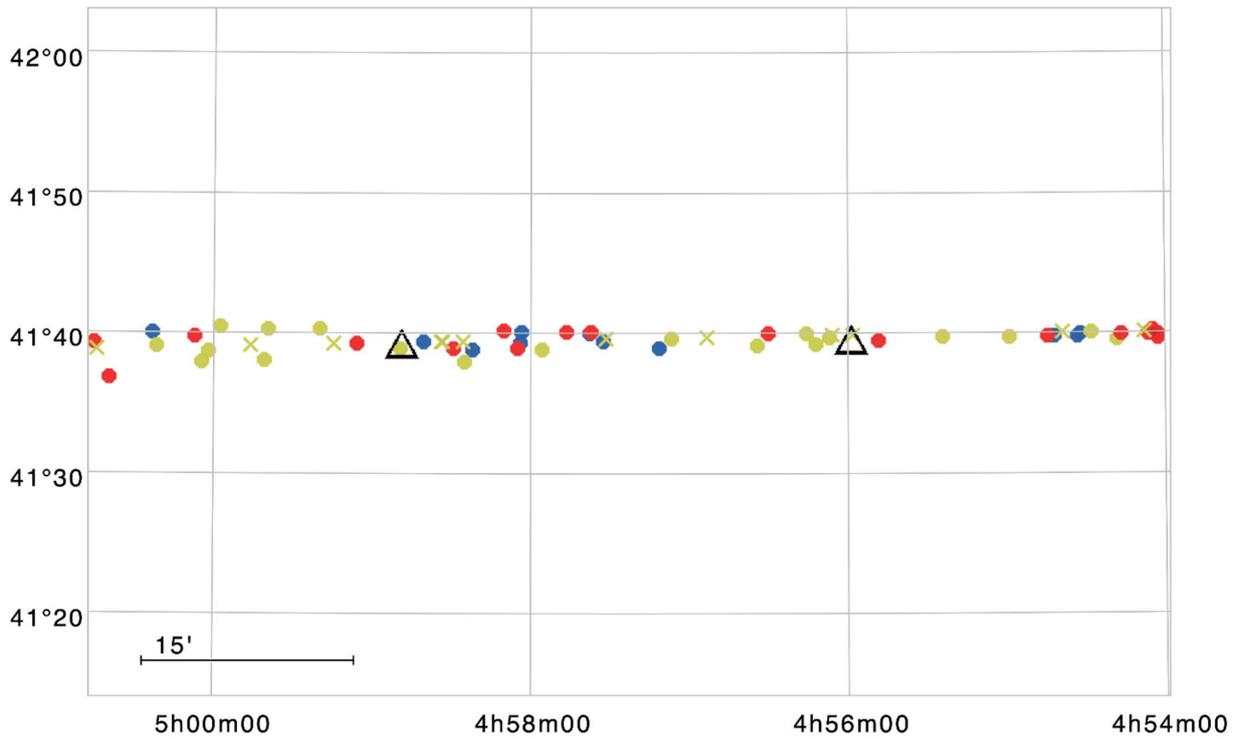

Figure 10. Modelled results of the observation made in 2014 August 27 (M5). X-axis shows RA and Y-axis DEC. Colour coding: Blue: SPs. Yellow: MPs. Red: BPs. Black triangles: start and end of the observed trail positions. Particles ejected towards the Sun are marked with crosses.

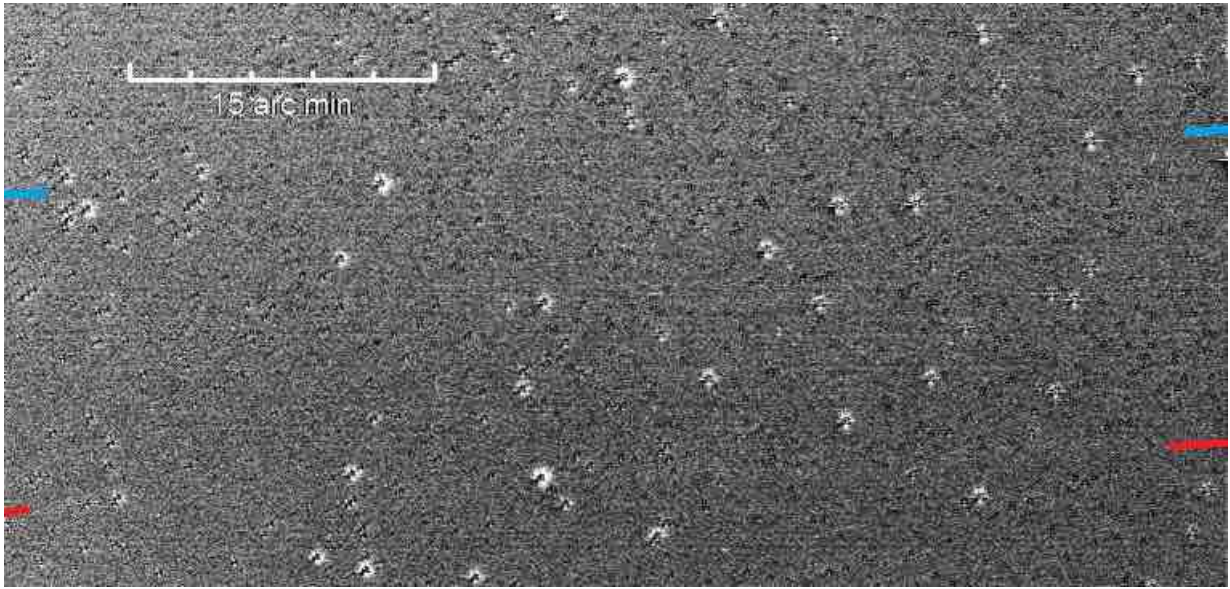

Figure 11. Observation made in 2014 August 27 (M5). Image subtraction. Lighter trail shown with the red markers is M5.

Observations were continued at the Auberry Sierra Remote Observatory and at the New Mexico Skies observatory in September 2014, when the comet itself was located on top of the dust trail as seen from Earth. All particle sizes were present during the observation with the comet itself (Figures 12-13).

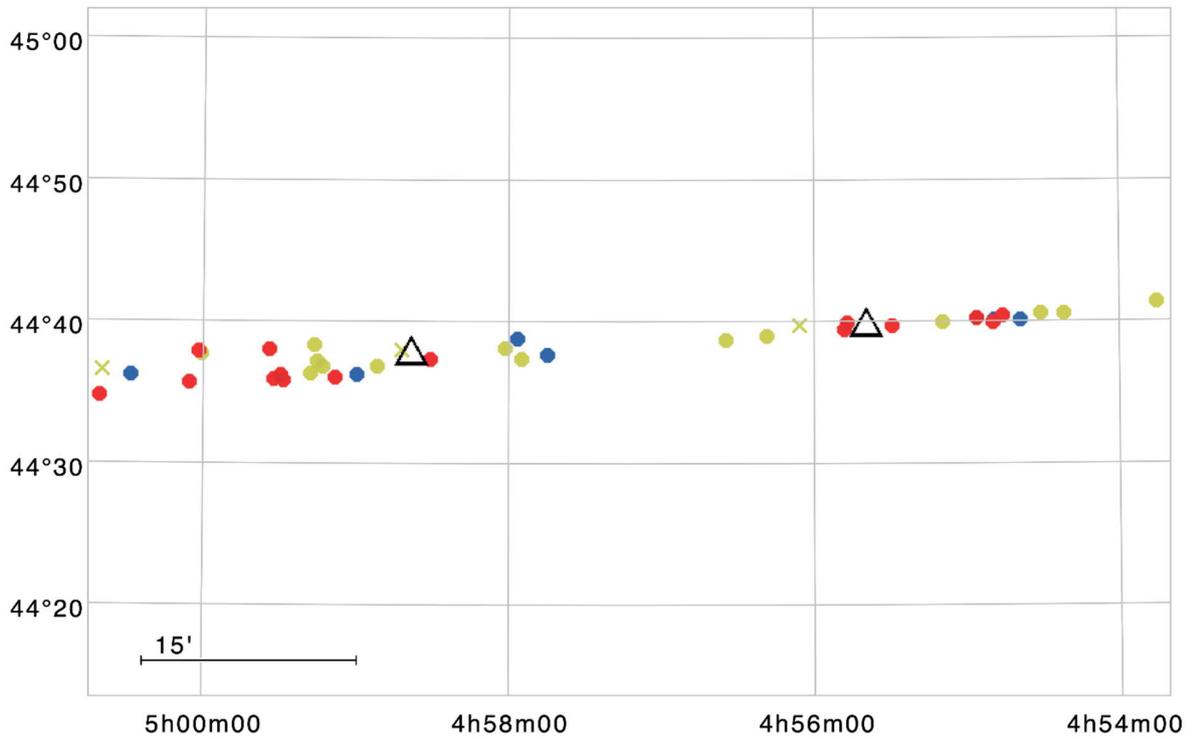

Figure 12. Modeling results vs observation made in 2014 September 16 (M6). X-axis shows RA and Y-axis DEC. Colour code of modeled particles: Blue: SPs. Yellow: MPs. Red: BPs. Black triangles: start and end observed positions of the trail. Particles ejected towards the Sun are marked with crosses.

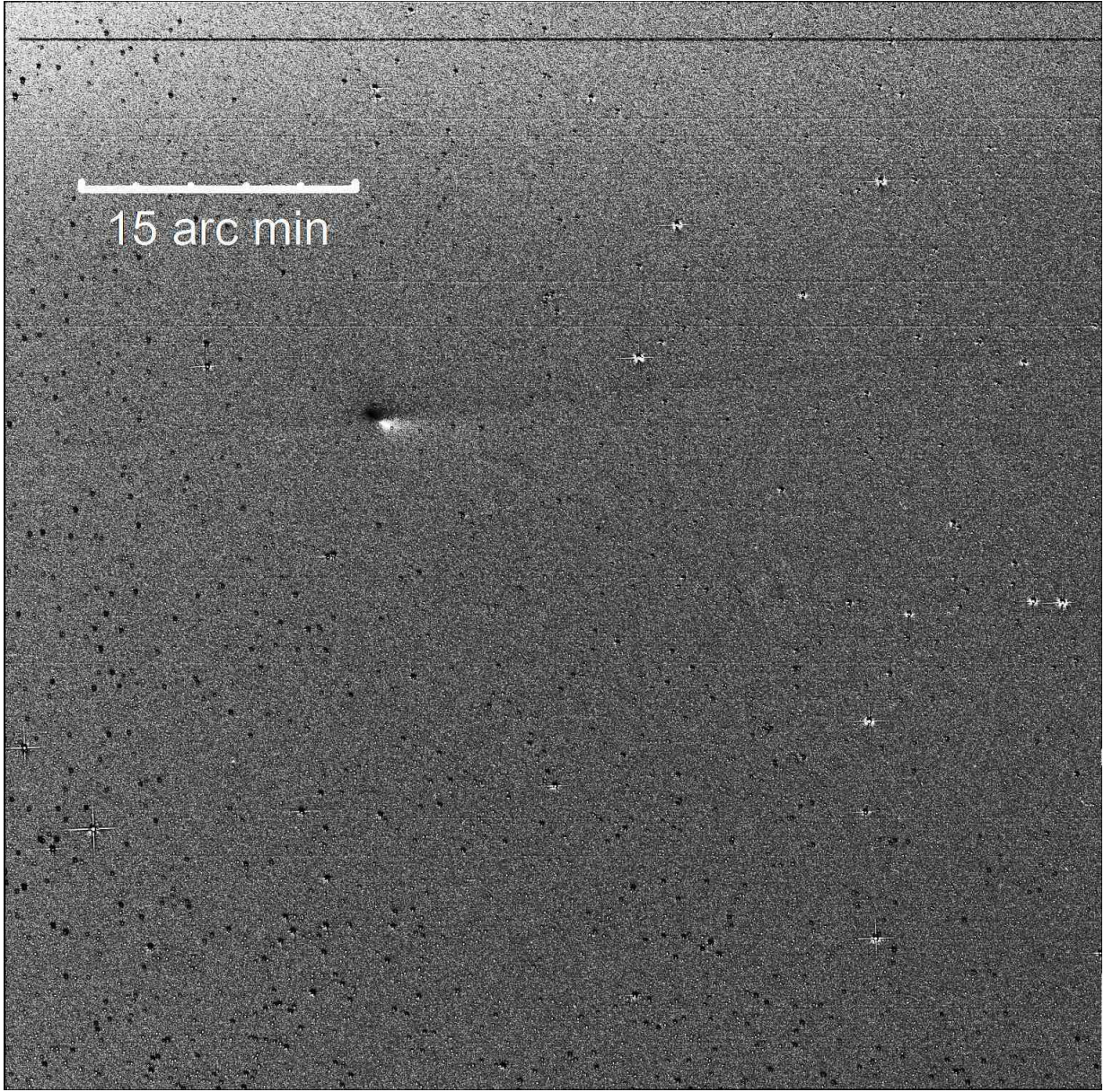

Figure 13. Image taken with the T24 telescope obtained when 17P/Holmes was aligned with the dust trail. The observation was made in 2014 September 16 (M6). Image subtraction. The darker trail is M6.

In September 2014, the Hankasalmi Observatory, Finland, also joined observations (Figures 14-15).

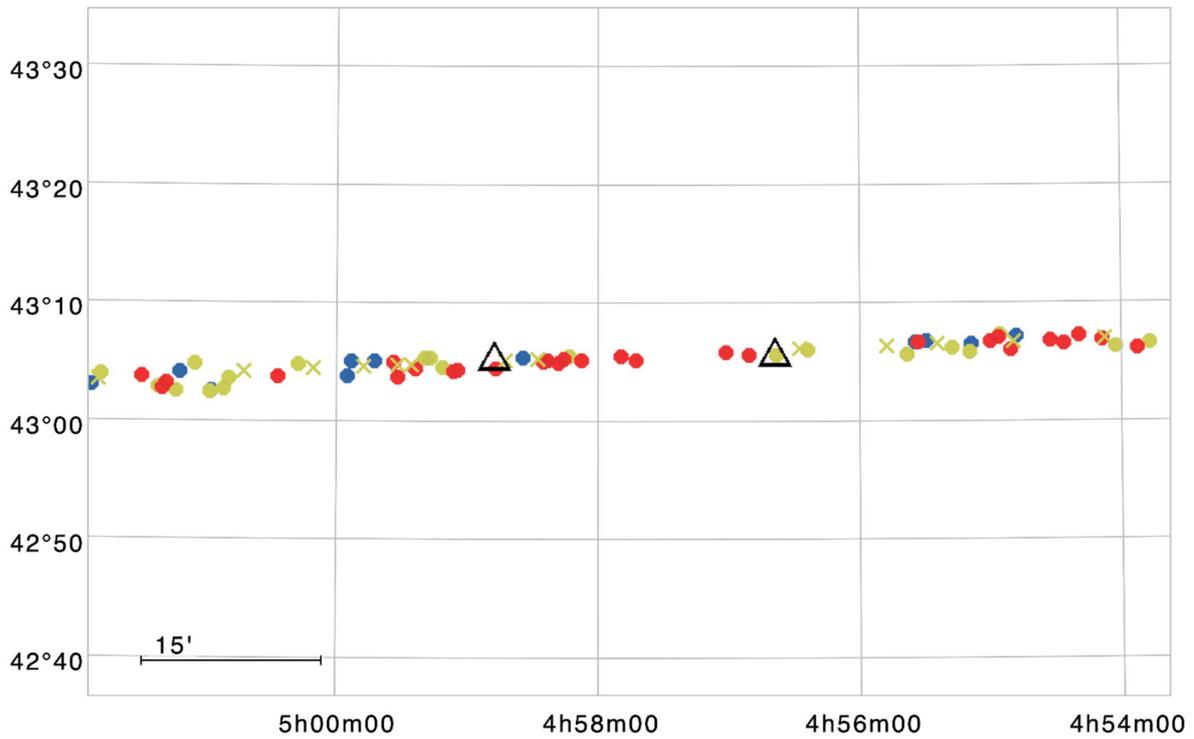

Figure 14. Modeling of the observation 2014 September 6 (M12). X-axis shows RA and Y-axis DEC. Blue: SPs. Yellow: MPs. Red: BPs. Black triangles: start and end observed positions of the trail. Particles ejected towards the Sun are marked with crosses.

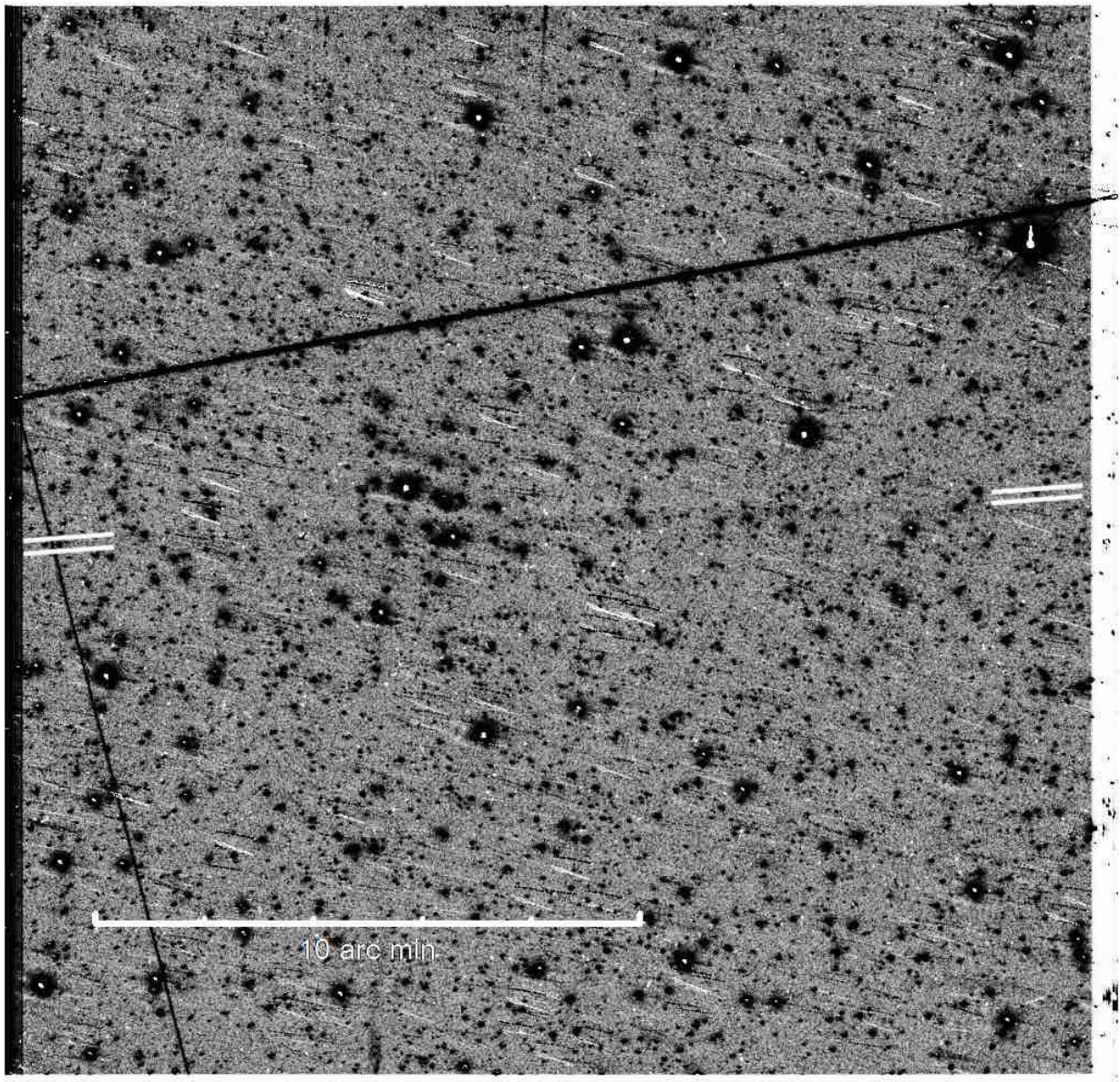

Figure 15. Observation made in 2014 September 6 (M12). Image subtraction. Lower trail is M12.

Starting from December 2014 the trail became brighter which made it possible to interpret the observations without image subtraction. All particle sizes were present in the vicinity of the 2007 outburst point. Additionally, mosaic images from the dust trail were made at the Hankasalmi observatory in February 2015 (Figures 16-21).

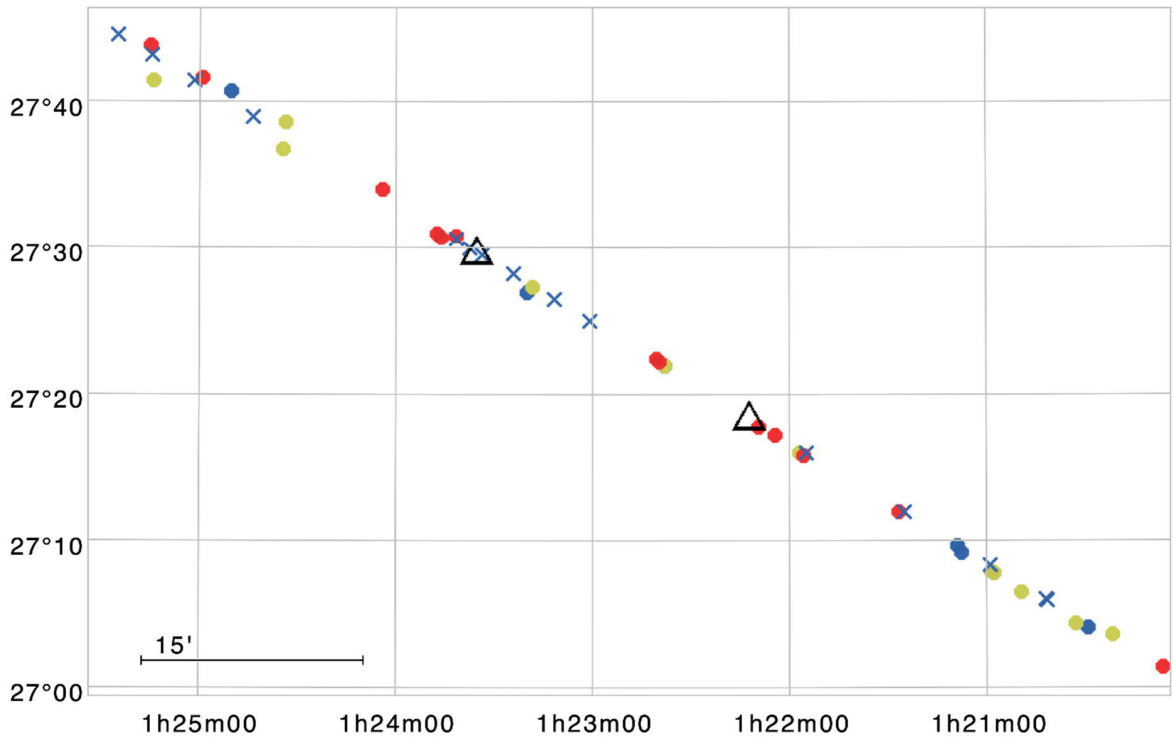

Figure 16. Modeling of observation 2015 February 14 (M13). X-axis shows RA and Y-axis DEC. Blue: SPs. Yellow: MPs. Red: BPs. Black triangles: start and end observed positions of the trail. Particles ejected towards the Sun are marked with crosses.

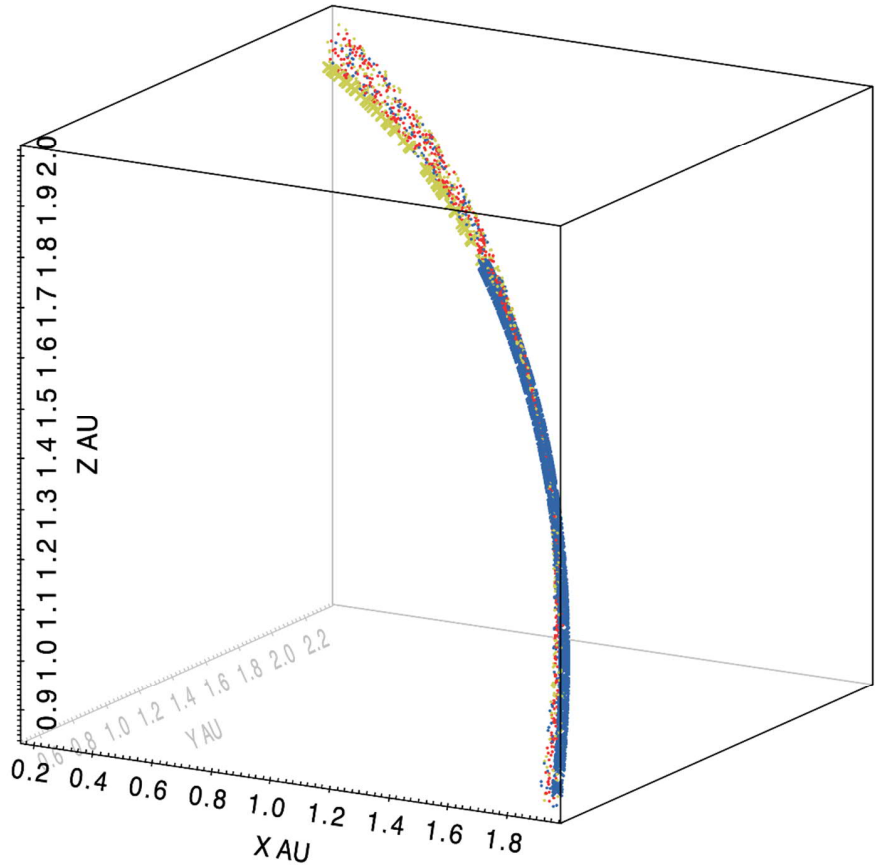

Figure 17. Modeling results vs observation made in the northern node in 2015 February 14 (M13). The particles are shown in the ICRF coordinates XYZ. Colour code stands for particle size. Blue: SPs. Yellow: MPs. Red: BPs.

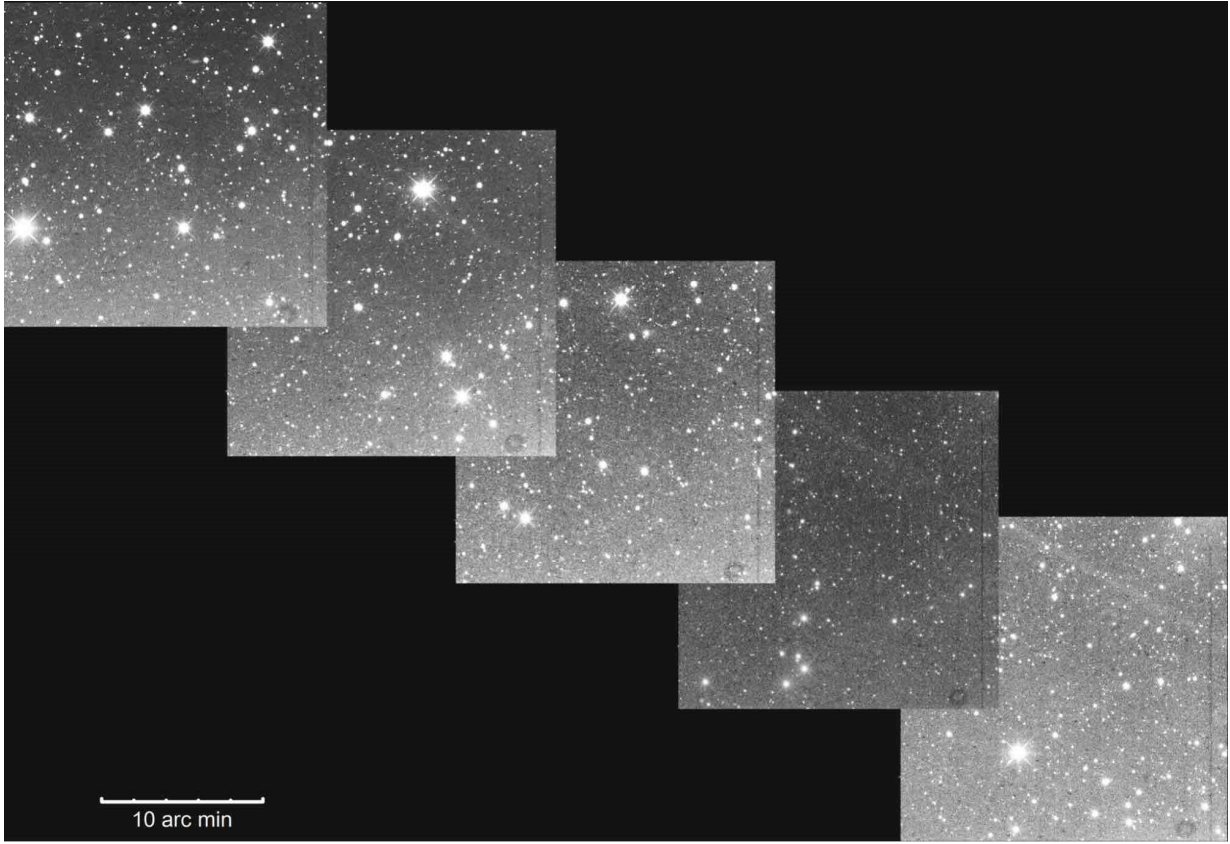

Figure 18. Observation made in 2015 February 14 (M13). No image subtraction was required.

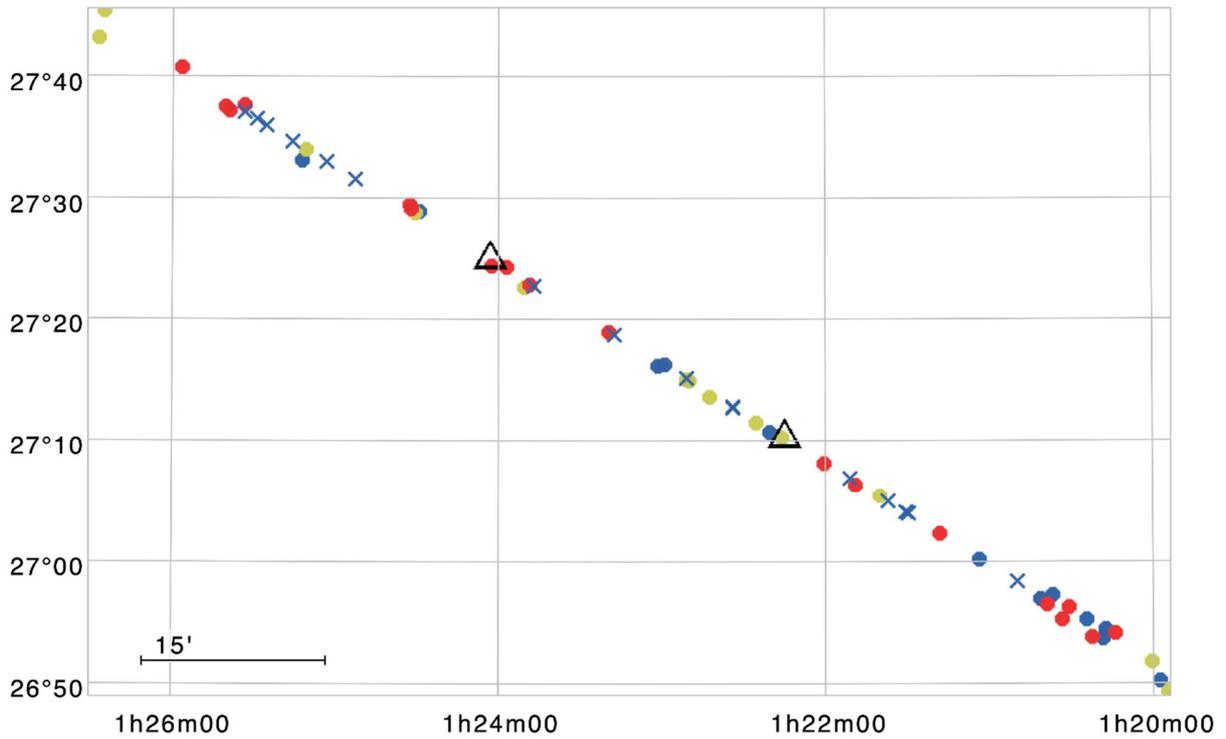

Figure 19. Modeling of the observation 2015 February 15 (M14). The X-axis shows RA and the Y-axis DEC. Blue: SPs. Yellow: MPs. Red: BPs. Black triangles: start and end observed positions of the trail. Particles ejected towards the Sun are marked with crosses.

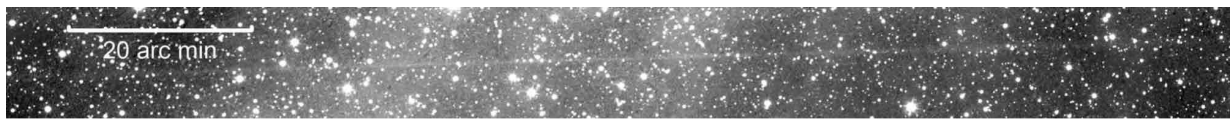

Figure 20. Observation of the trail as seen in 2015 February 15 (M14). Without image subtraction.

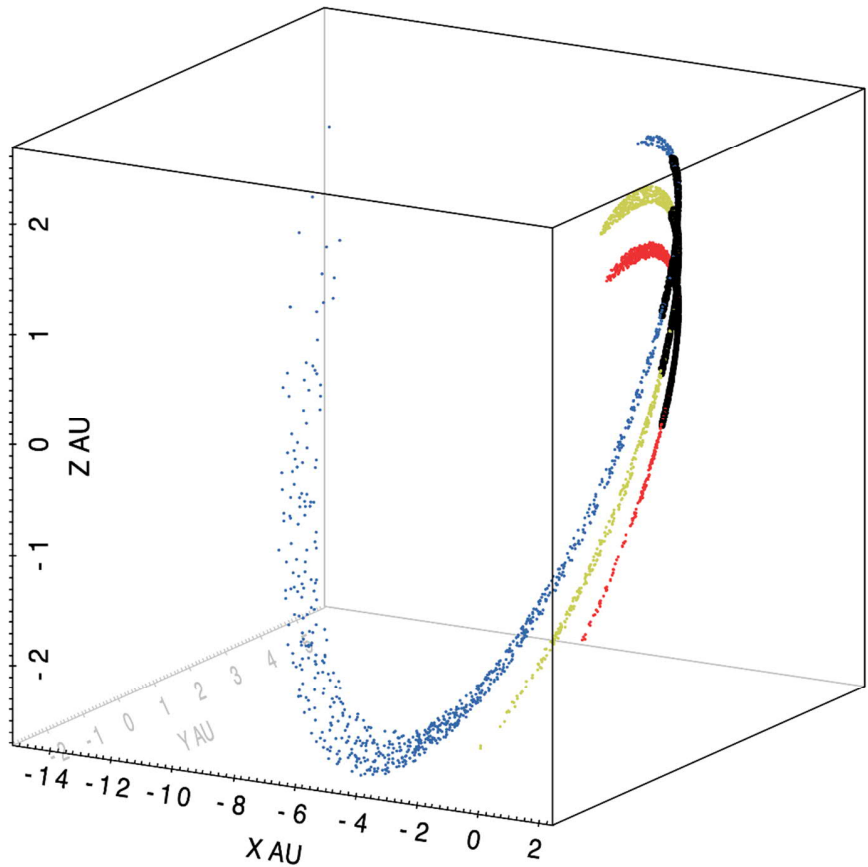

Figure 21. Modeling results for the time of the observation made in 2015 February 15 (M14). Here we show a complete modeled dust trail. The 40° RA sections of the trail are coloured here by black. The particles are shown in the ICRF coordinates XYZ. Blue: SPs. Yellow: MPs. Red: BPs. In order to fully demonstrate all particle populations, we have applied offset Z = 0.5 to the blue particles and offset Z = -0.5 to the red particles.

August 2015 observations from Auberry showed all particle sizes (Figures 22-23).

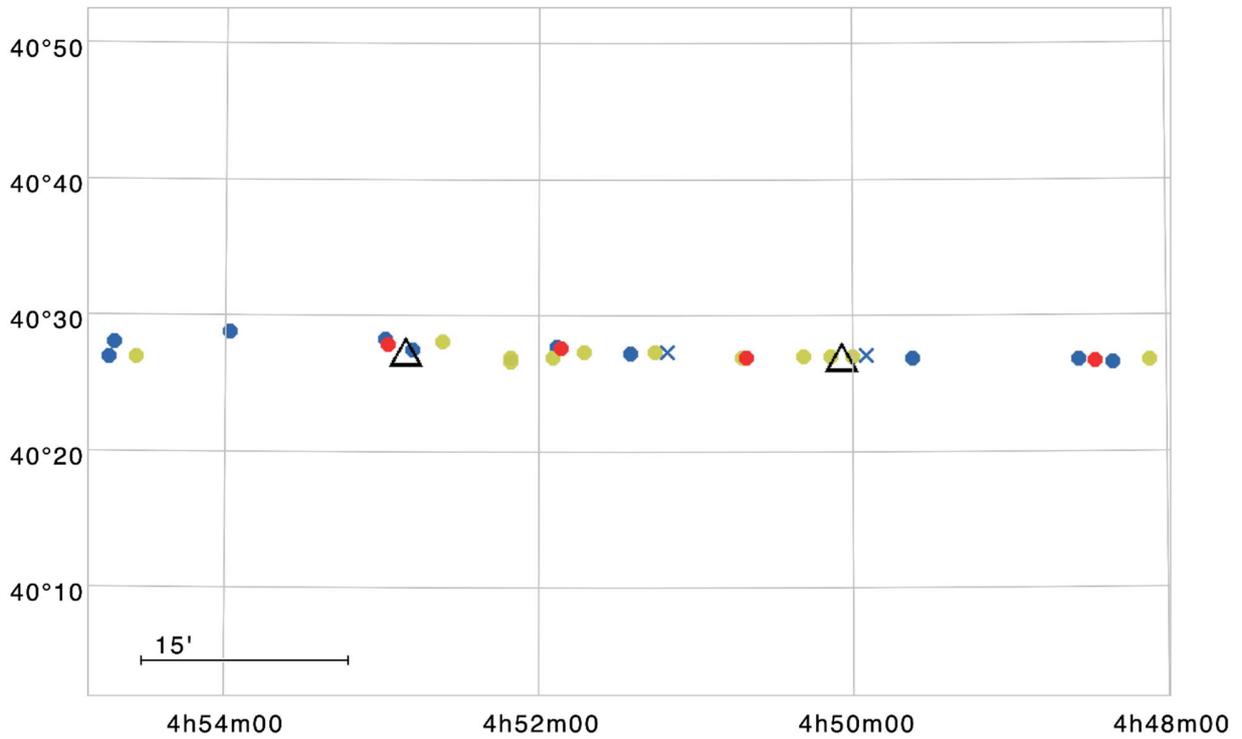

Figure 22. Modeling vs. observation made in 2015 August 18 (M10). The X-axis shows RA and the Y-axis DEC. Blue: SPs. Yellow: MPs. Red: BPs. Black triangles: observed start and end positions of the trail. Particles ejected towards the Sun are marked with crosses.

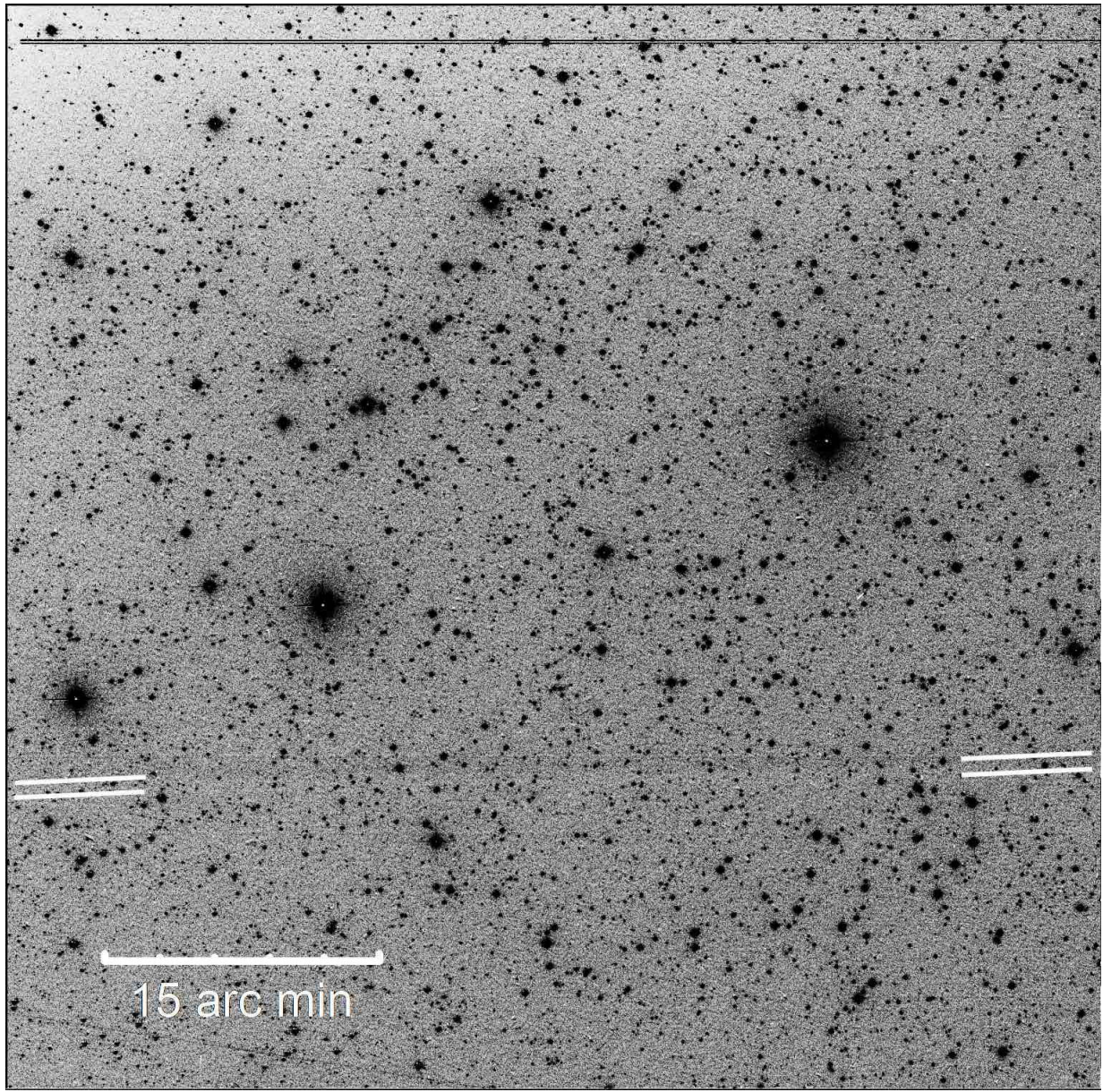

Figure 23. Observation 2015 August 18 (M10). Image subtraction. Darker trail is M10.

Last observations of the trail were made in October 2015 from Auberry. Then only small and medium sized particles were present at the 2007 outburst point. The density of medium sized particles was decreasing by a factor of ~0.7 and the small particles started dominating the 2007 outburst point (Figures 24-25).

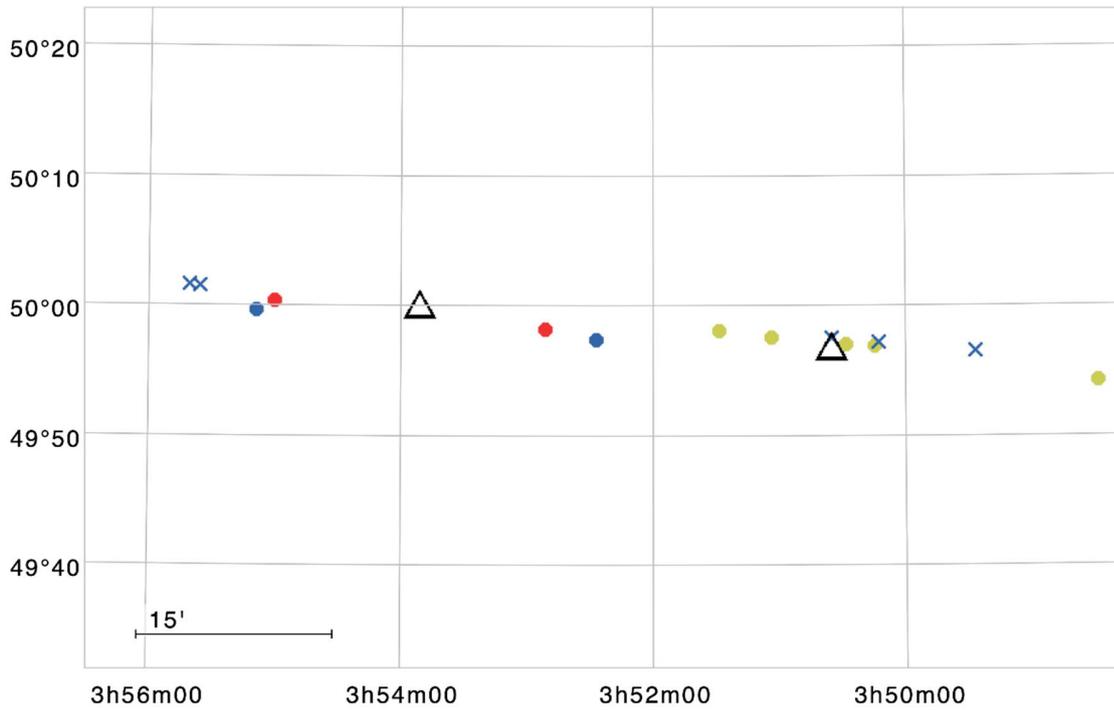

Figure 24. Modeling of observation 2015 October 24 (M11). The X axis shows RA and the Y axis DEC. Blue: SPs. Yellow: MPs. Red: BPs. Black triangles: start and end observed positions of the trail. Particles ejected towards the Sun are marked with crosses.

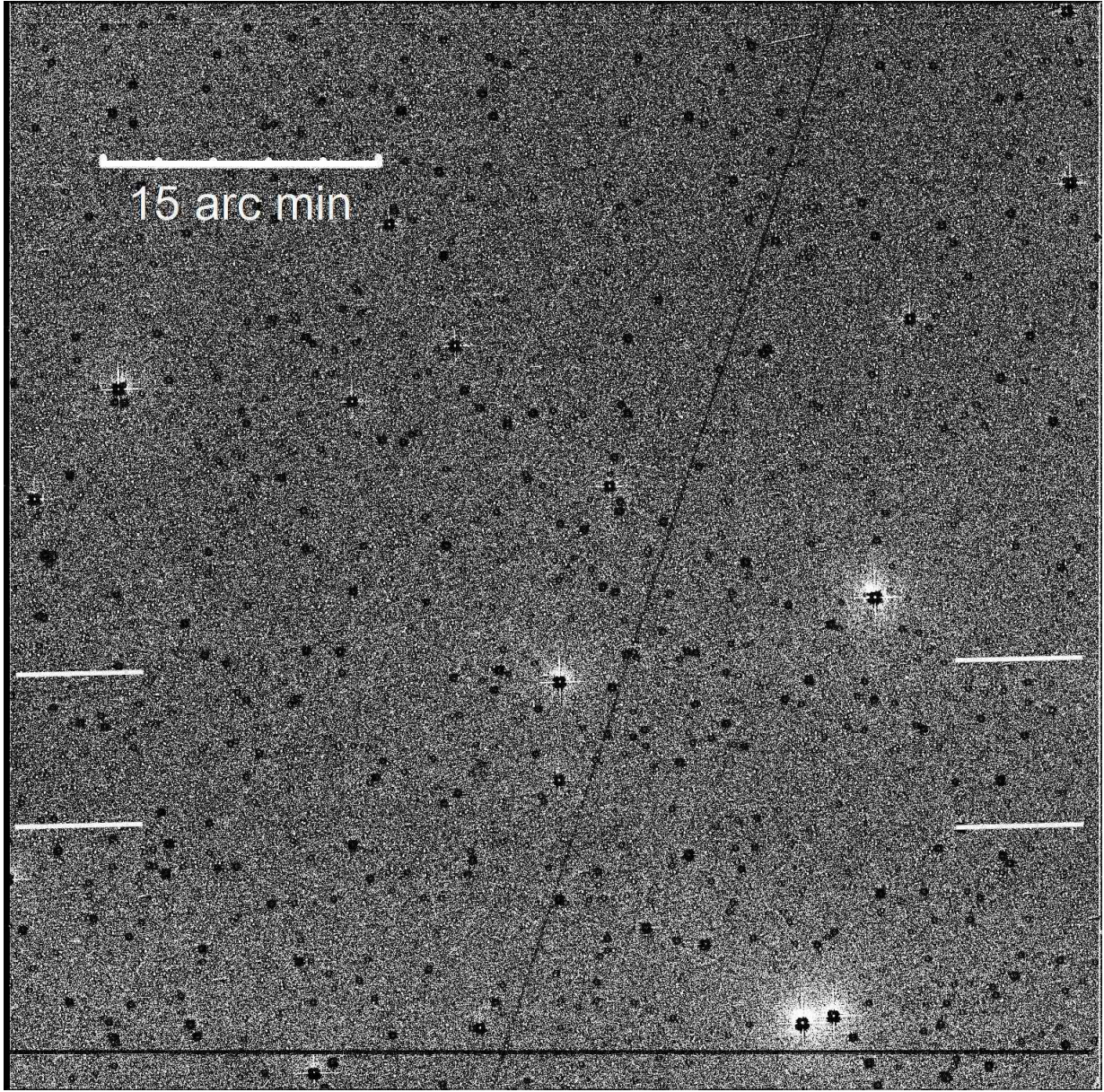

Figure 25. Observation 2015 October 24 (M11). Image subtraction. The upper trail is M11.

4.4. Observing campaign 2020 - 2021

The search of the trail was initiated in September 2020 from the New Mexico Skies observatory. The small particles had not yet traveled to the vicinity of the 2007 outburst point. However, the big and medium sized particles were present there. The search continued into October 2020. While the small particles had not yet arrived, the medium and big sized particles were present with the increased density. No trail was visible even when using image subtraction technique.

A continued search of the trail was made again in March 2021 from New Mexico Skies observatory. The small particles started populating the 2007 outburst point, but not with notable density. The big and medium sized particles were present with even higher density compared to September 2020 and October 2020. The dust trail was likely low in the sky (~20-30° in altitude) since it was not visible with image subtraction technique in the observations.

Physically, the trail was even further away from the original 2007 outburst point than during the successful observation time, in February 2015, and is expected to remain as such in February 2022 predictions.

In August 2021, all particles were present at the 2007 outburst point, but the density of the small particles was not yet high (see supporting Figures in Supplementary Material). The comet itself was near the trail in August and at the beginning of September 2021. The comet has crossed the narrowest part of the trail in the middle of August 2021. In Supplementary Material the comet is plotted on top of the modeled trail section for 2021 September 6. The convergence point movement in the sky is also shown.

4.5. Predictions for 2022

The density of small particles is not expected to increase significantly until well into 2022 (Figures 26-28). The physical dust trail is predicted to move towards the original 2007 outburst point. The width of the trail seen from Earth in February and March of 2022 will be comparatively similar to the February 2015 observations. The particle density is expected to be comparatively similar as well. All particle sizes will be present in the trail near the 2007 outburst point.

The brightness of the February 2015 trail was bigger than our spherically symmetrical or towards the Sun outburst modelling achieve. Alternative explanation of the increase in actual brightness of the trail could be ejection of additional material from the comet relative to what is assumed in our model, e.g., during or after to the time of the 2007 outburst.

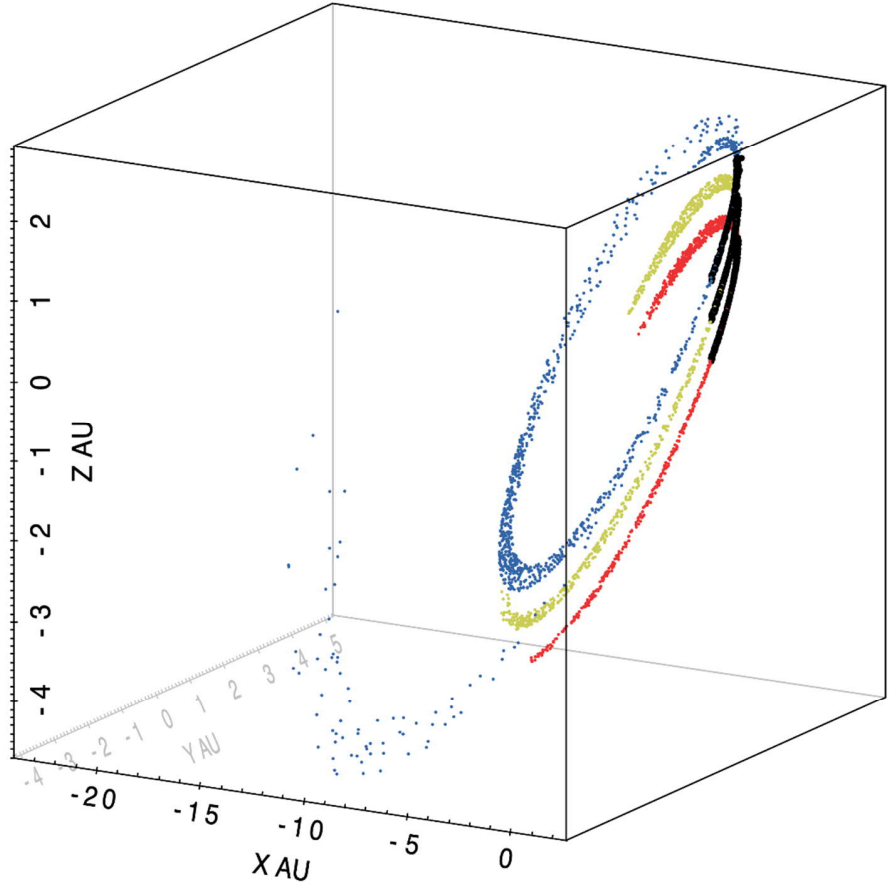

Figure 26. Modeling of the predicted trail for February 2022 (2022-02-15T12:00:00) (F7). The section highlighted is modeled in more detail near the 2007 outburst point. Blue: SPs. Yellow: MPs. Red: BPs. Black circles: 2007 outburst point centered 40° RA window. Particles are shown in the ICRF coordinates XYZ. In order to fully demonstrate all particle populations, we have applied offset Z = 0.5 to the blue particles and offset Z = -0.5 to the red particles.

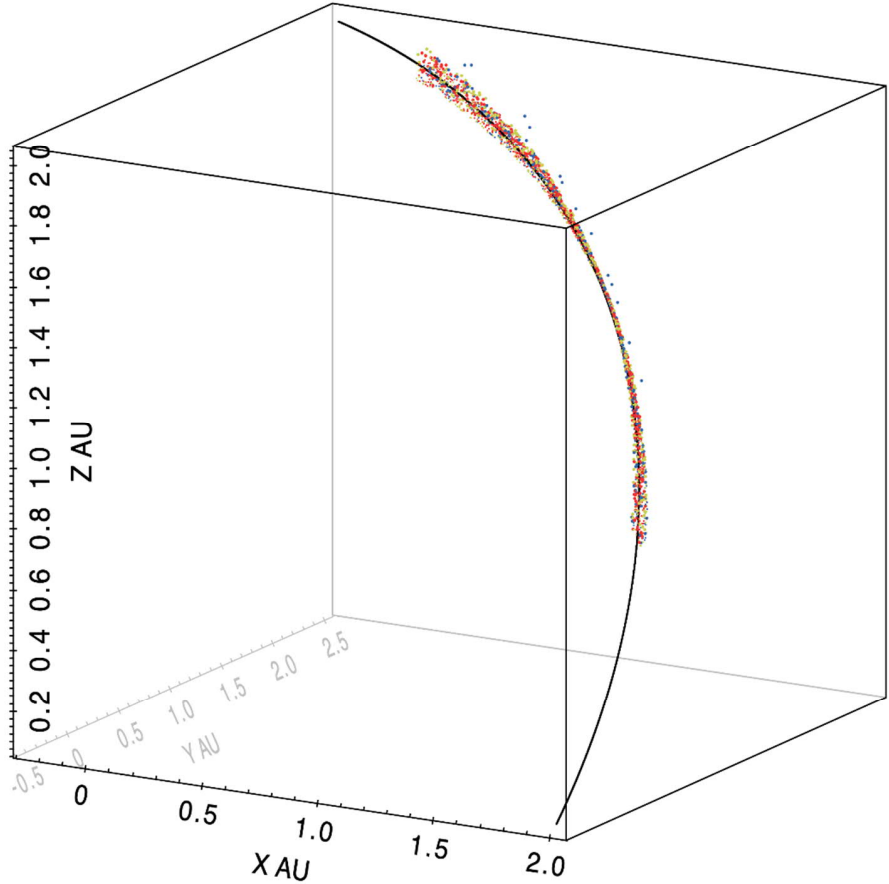

Figure 27. Prediction of the dust trail in February 2022 (2022-02-15T12:00:00) (F7) near the 2007 outburst point. Marked in the figure is the 17P/Holmes orbit at 2007 outburst event, the modeled February 2015 trail (2015-02-14T12:00:00) (M13) and 0.01 AU further away the modeled February 2022 trail. The particles are shown in the ICRF coordinates XYZ.

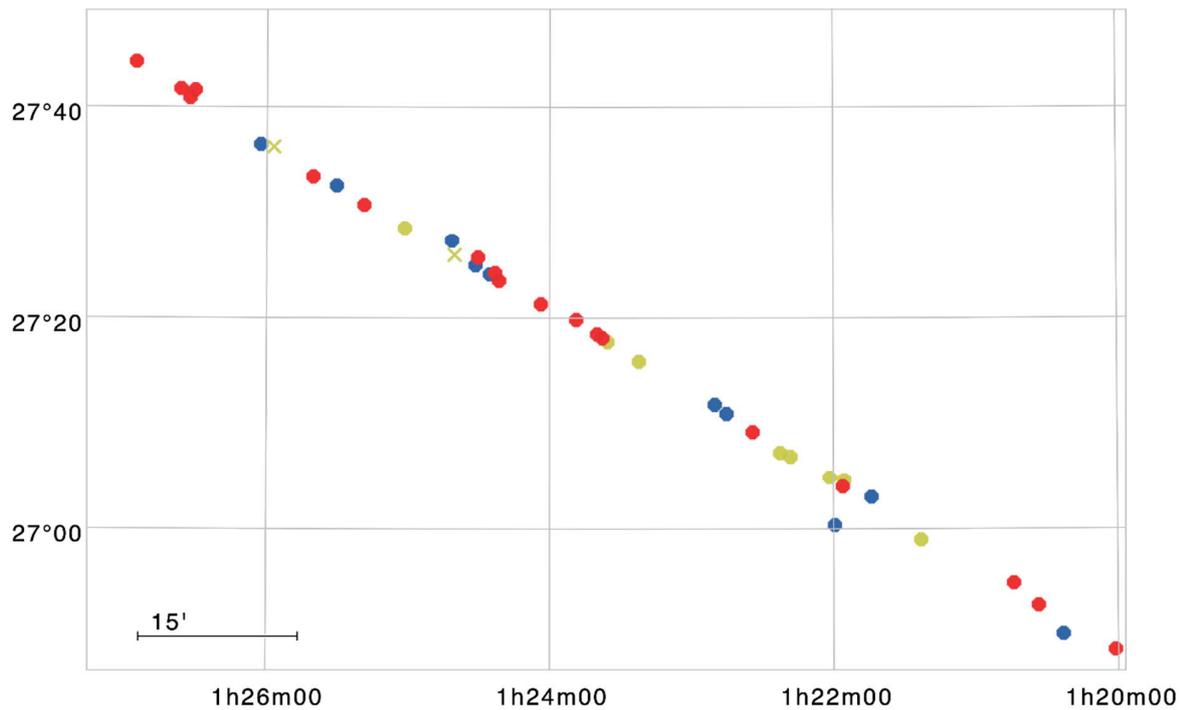

Figure 28. Modeling of the dust trail in February 2022 (2022-02-15T12:00:00) (F7). The X-axis shows RA and the Y-axis DEC. Blue: SPs. Yellow: MPs. Red: BPs. Particles ejected towards the Sun are marked with crosses.

In August 2022, the density of big particles is starting to decrease. The trail takes up the space of considerably more than a full orbit length and includes the dispersed tail in 2022. In Supplementary Material we show the visualization of hourglass pattern in both, near the 2007 outburst point and at the southern orbit node. The modeling of the dust trail in August 2022 are shown in Figure 29.

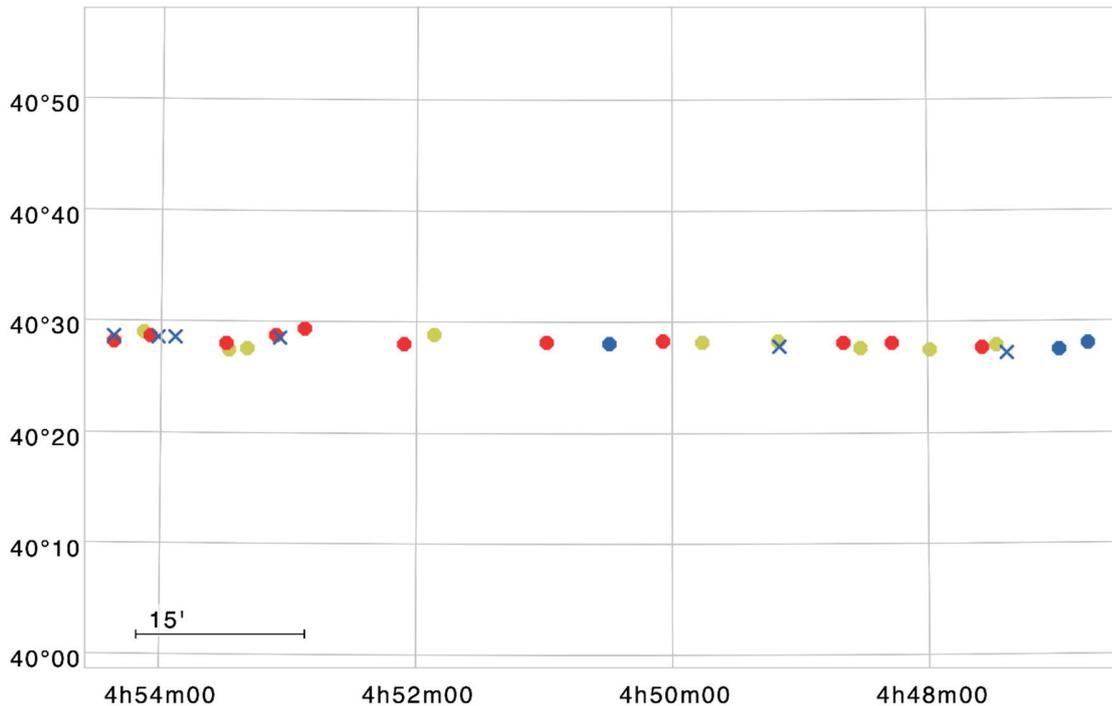

Figure 29. Modeling of the dust trail in August 2022 (2022-08-18T12:00:00) (F9). The X axis shows RA and the Y axis DEC. Blue: SPs. Yellow: MPs. Red: BPs. Particles ejected towards the Sun are marked with crosses.

The dust trail should be within the reach for even modest telescopes in 2022. However, image subtraction techniques are needed when observing with a small telescope (Alard, 2000). We estimate that the brightness of the two-revolution dust trail is nearly similar to or brighter than that during the 2013 observations of the dust trail in the far side node, when the surface brightness was 26.8 mag/arcsec$^2$ and the phase angle was 15.56° (Lyytinen et al., 2013b). The brightness value was also measured, when dust trail was directly observed during 2015 observations near the 2007 outburst point in the northern hemisphere (Lyytinen et al., 2015). We further estimate that the two-revolution dust trail is less bright, than that surface brightness of 25 mag/arcsec$^2$ when the phase angle was 21.45°. The brightness estimates provided here are based on comparing the modeled trail widths and the densities for the entire range of modeled particles. The phase angles difference and therefore changed albedo causes 20% brightness increase for February 2013 observations compared with February 2015 observations (Kolokolova et al., 2004). There is however uncertainty of the reason for the brightness increase in February 2015 trail (as discussed above) and that requires further investigations for predictions.

When accounting for the solar radiation pressure effects and modeled particle size distribution, the measurements we made by direct observing are consistent with the modeling results. The trail position is in good agreement with the modelling and the narrowest points of the trail are in good agreement with the spherically symmetric particle ejection model.

There can be other non-gravitational and non-regular radiation pressure effects acting on the particles, such as seasonal type radiation effects, which can slow down the particles even more than our modeling can account for.

We have not directly observed or modeled such effects at this juncture, because the behaviour of the particles in the dust trail suggests that they are influenced mainly by solar radiation pressure and Jupiter's gravitational effects. Future observations are expected to shed more light on the magnitude of such secondary effects.

## 5. FUTURE OBSERVATIONS

Our prediction with the particles' calculated coordinates of the centre of the trail section nearest to the 2007 outburst point for years 2021 and 2022 are shown in Supplementary Material.

Direct observations in the vicinity of the hourglass centre and the dust trail in 2022 and later may likely provide further information about the following:

- symmetry conditions and exact mechanism of the 2007 outburst,
- possible dispersion of material in interplanetary space,
- characterizing non-regular radiation pressure effects, such as seasonal type radiation effects to the particles, as it requires longer time span of observations than a few revolutions,
- particle size distribution evolution in the dust trail.

Observations can aid in calculating the following parameters:

- the extent of the phenomena measured in time,
- position of the dust trail,
- width of the trail and width of the hourglass centre,
- the brightness evolution of the dust trail in time.

The density of particles in the trail will be lower, density for all particles in February 2022 prediction is ~0.7 times the density in February 2015 modeling results (Figure S1, S2). Phase angle was 21.45° in February 2015 and 21.25° in February 2022. Albedo of particles is dependent on the phase angle. Phase angles are on all occasions in this study between 12° and 25°. Albedo difference is here below 10% in the least squares fit to the data presented by Kolokolova et al. (2004).

Upon discovery on 1892 November 6 by Edwin Holmes, comet 17P/Holmes also underwent an outburst at that time that led to the discovery of the comet (Hsieh et al., 2010). It could be beneficial to model that outburst in more detail as well as to examine if some effects of the 1892 outburst could be still observable. In a longer interval, non-regular radiation effects can become more prominent.

If the previously generated dust trails are visible in the future, they will be deflected even farther away. It is not clear if they are sufficiently bright for direct observations.

According to our model, there will be no further loss of the particles after the initial mass loss from the trail has occurred. If there is a loss, vanishing or break up of the particles, not accounted for in our model, the phenomenon might be dimmer than expected. In the previous observations we have not identified any substantial brightness decrease that has resulted from the loss of material. This is however difficult to measure, because the outburst mechanism is not known in such detail, and it is difficult to determine optical depth test particles that are consistent with both model and observations.

In 2022 the dust will be observable in visible light. While dust particles are not bright in the near-infrared spectrum, they could occasionally also be observable, where they will become brighter. Dust particles will be observable also in the mid-infrared spectrum.

The interaction of light and particles in the comet's trail is currently a subject of our investigation and the results will be reported in a separate later publication.

## 6. CONCLUSIONS

This paper describes a comprehensive particle model that was built upon the previous model developed by Lyytinen (1999), Lyytinen and Van Flandern (2000), and Lyytinen et al. (2001). The improved model includes multiparticle Monte Carlo modeling and was enhanced by including the solar radiation pressure effects, gravitational disturbance caused by Venus, Earth and Moon, Mars, Jupiter and Saturn, and gravitational interaction of the dust particles with their parent comet. We use this model to describe the dust trail evolution of the comet 17P/Holmes for a full time period since the spectacular outburst in October 2007.

For the first time, the hourglass pattern of a comet trail has been observed, and modeling was used to explain it. We found that spherical symmetry of the ejected particles is responsible for producing the hourglass pattern (vs. a purely theoretical case when all particles are ejected towards the sun).

The spherically symmetric outburst model is not able to explain the concentration of particles implied by the sharp increase in brightness of the trail near the outburst location in February 2015 observations. When we add the towards the Sun model, it brings small particles directly to the centreline of the trail, but not sufficiently close.

When comparing the February 2013 model vs. observations, the spherically symmetric model gives a better agreement. Towards the Sun ejected particles concentrate on the centreline of the trail, which results in an overly bright trail compared to what was observed.

In all cases where the hourglass pattern of the trail was observed, it also resulted from our model with spherically symmetric modelling of the outburst. Towards the Sun ejected particles do not contribute to the creation of a hourglass pattern, instead they produce a sharp concentrated line of particles on the trail centreline.

To support our model, we have conducted ground-based telescopic observations of the dust trail when the particles converge in any of the two common nodes (in the near-side common node or the far-side common node). The observations were done in visible light and span for a significant period, from 2013 to 2015. The new improved model was successfully validated by directly comparing the dust trail particle observations with the modeled particle distributions during a time span extending for several years. The modelled trail position, width and narrowest point determination of the trail were found to be in excellent agreement with the observations.

Further to this, the improved model was used to constrain future evolution of the dust trail by providing predictions for the trail position and visibility. At present, the two-revolution dust trail is twice as dispersed temporally, and at the same time it is dimmer than when the particles had travelled one revolution around the Sun. However, near the 2007 outburst point, the apparent observable radial dispersion is not significantly larger.

The dust trail is expected to be deflected further away by approximately 0.01 AU in March of 2022, mainly due to the gravitational disturbance of Jupiter, and solar radiation pressure to some extent. The distance varies slightly, and the trail was located a few hundredths of AU farther away in February of 2021.

We predict that the evolved dust trail of the comet 17P/Holmes should be visible with even modest telescopes in 2022. Observing the dust trail in 2022, or later, will provide even more insight into understanding this phenomenon. Continuous future observations will also enable further modelling development since long-term ground truth information is essential for validating and improving the models.

ACKNOWLEDGEMENTS

This work was supported, in part, by the Academy of Finland project no. 325806 (PlanetS). The authors express deep gratitude and dedication to Esko Lyytinen, in particular, for initiating this research and for putting in place effective collaboration under the umbrella of the Ursa Astronomical Association and the Finnish Fireball Network. Intense collaborative work with Esko allowed us to provide a comprehensive representation of the ideas earlier discussed with Esko in the form of personal communications. We are grateful to Pekka Lehtikoski for his contribution to the programming of the mathematical model. We thank Jérémie Vaubaillon for insightful comments and discussion, which helped us to improve the earlier version of this paper. We thank Jorma Ryske for his enthusiastic observations of the trail in February and March 2022 and for confirming the observability and position of the trail as described and predicted in this study. This research made use of TOPCAT for visualization and figures (Taylor, 2005).

Data Availability

The data underlying this study are included in the article and supporting information. TOPCAT files containing the output modeling data used for visualization can be shared on a reasonable request to the corresponding author.

Appendix. List of Symbols

$\beta$ = ratio of radiation pressure to gravity
$\mu$ = gravitational parameter of the Sun
$\mu'$ = Solar radiation pressure affected gravitational parameter of the Sun effective to particle
$v_{ex}$ = ejection velocity x component
$v_{ey}$ = ejection velocity y component
$v_{ez}$ = ejection velocity z component
$v_{0x}$ = Particle's velocity at the start location (x component)
$v_{0y}$ = Particle's velocity at the start location (y component)
$v_{0z}$ = Particle's velocity at the start location (z component)
$v_x$ = Particle's velocity x component after ejection speed is added
$v_y$ = Particle's velocity y component after ejection speed is added
$v_z$ = Particle's velocity z component after ejection speed is added

$x_{\text{earth}}$ = Earth's location x component in the International Celestial Reference Frame (ICRF/J2000) standard celestial reference system

$y_{earth}$ = Earth's location y component in the International Celestial Reference Frame (ICRF/J2000) standard celestial reference system

$z_{earth}$ = Earth's location z component in the International Celestial Reference Frame (ICRF/J2000) standard celestial reference system

$x_p$ = Particle's location x component in the International Celestial Reference Frame (ICRF/J2000) standard celestial reference system

$y_p$ = Particle's location y component in the International Celestial Reference Frame (ICRF/J2000) standard celestial reference system

$z_p$ = Particle's location z component in the International Celestial Reference Frame (ICRF/J2000) standard celestial reference system

$x_g$ = Geocentric x component of the particle

$y_g$ = Geocentric y component of the particle

$z_g$ = Geocentric z component of the particle

$\alpha$ = Declination

$\theta$ = Right ascension

# Supplementary Material for:
# Evolution of the Dust Trail of Comet 17P/Holmes

Maria Gritsevich[1,2,3,4], Markku Nissinen[2], Arto Oksanen[5], Jari Suomela[6], Elizabeth A. Silber[7,8]

1. Finnish Geospatial Research Institute (FGI), Vuorimiehentie 5, FI-02150 Espoo, Finland
2. Finnish Fireball Network, Ursa Astronomical Association, Kopernikuksentie 1, FI-00130 Helsinki, Finland
3. Department of Physics, University of Helsinki, Gustaf Hällströmin katu 2a, P.O. Box 64, FI-00014 Helsinki, Finland
4. Institute of Physics and Technology, Ural Federal University, street of Peace 19, 620002 Ekaterinburg
5. Hankasalmi observatory, Jyväskylän Sirius ry, Verkkoniementie 30, 40950 Muurame, Finland
6. Clayhole observatory, Jokela, Finland
7. Department of Earth Sciences, Western University, London, ON, N6A 5B7, Canada
8. The Institute for Earth and Space Exploration, Western University, London, ON, N6A 3K7, Canada

**Table S1:** List of observations. Observations: iTelescope Siding Spring Observatory, Mayhill and Auberry Observatories (T30, T24, T11, T21). Hankasalmi Observatory observations (H). The measurement accuracy from the observation images is 30 arc seconds for image subtracted images and 10 arc seconds for non-image subtracted images.

| JD (UT) | RA (trail start) | DEC (trail start) | RA (trail end) | DEC (trail end) | Telescope | Measurement number |
|---|---|---|---|---|---|---|
| 2456341.1895139 | 241.3250 | -40.1731 | 242.2375 | -40.2567 | T30 | M2 |
| 2456343.2232523 | 241.3292 | -40.3350 | 241.9792 | -40.4247 | T30 | |
| 2456514.9011458 | 207.3875 | -31.3539 | 207.9375 | -31.6172 | T30 | |
| 2456515.9078588 | 207.3750 | -31.2531 | 207.9333 | -31.5194 | T30 | |
| 2456529.0698148 | 208.8708 | -30.8381 | 209.4250 | -31.0658 | T30 | M3 |
| 2456529.9174074 | 208.8667 | -30.7392 | 209.4167 | -30.9900 | T30 | |
| 2456532.1898727 | 209.1417 | -30.6881 | 209.4083 | -30.8133 | T30 | |
| 2456534.9117708 | 209.2083 | -30.4964 | 209.6625 | -30.7017 | T30 | |
| 2456700.2390394 | 240.0083 | -39.4950 | 240.6125 | -39.5792 | T30 | M4 |
| 2456701.2430208 | 240.0125 | -39.5772 | 240.6167 | -39.6586 | T30 | |
| 2456895.7358102 | 73.6875 | 41.5211 | 74.3917 | 41.5139 | T24 | |
| 2456896.9660417 | 73.9917 | 41.6511 | 74.7000 | 41.6472 | T24 | M5 |
| 2456904.0601273 | 70.0542 | 42.6667 | 70.8708 | 42.6922 | T11 | |
| 2456904.9456134 | 70.0500 | 42.8278 | 70.8458 | 42.8294 | T11 | |
| 2456906.5282176 | 74.2083 | 43.0142 | 74.7458 | 43.0086 | H | M12 |
| 2456916.0087037 | 73.9333 | 44.5050 | 74.6750 | 44.4653 | T24 | |
| 2456916.9346065 | 73.9208 | 44.6492 | 74.6667 | 44.6164 | T24 | M6 |
| 2456916.9876620 | 74.1917 | 44.6589 | 74.9417 | 44.6108 | T24 | |
| 2456917.9345602 | 74.1833 | 44.7997 | 74.9333 | 44.7564 | T24 | |
| 2456917.9946644 | 74.1833 | 44.8228 | 74.9333 | 44.7769 | T24 | |
| 2456923.1607870 | 72.7875 | 45.6808 | 73.5500 | 45.6408 | T24 | |
| 2456924.0018981 | 72.7792 | 45.8375 | 73.5375 | 45.7981 | T24 | |
| 2456925.8736227 | 71.5125 | 46.1844 | 72.2792 | 46.1508 | T24 | |
| 2457017.7527083 | 16.6375 | 34.2192 | 17.2625 | 34.5447 | T24 | M7 |
| 2457037.6497685 | 15.9250 | 29.8197 | 16.5208 | 30.1797 | T24 | |
| 2457042.7076505 | 16.8792 | 29.4358 | 17.4750 | 29.7800 | T24 | M8 |
| 2457043.5607176 | 16.8875 | 29.2519 | 17.4833 | 29.5978 | T24 | M9 |
| 2457064.3084722 | 20.5583 | 27.8519 | 20.7917 | 27.9808 | H | |
| 2457064.6718403 | 20.4958 | 27.7619 | 21.0167 | 28.0517 | T24 | |
| 2457066.3566319 | 19.9625 | 27.2192 | 20.4083 | 27.4697 | H | |
| 2457066.6683102 | 20.8083 | 27.6419 | 20.3417 | 27.3822 | T24 | M1 |
| 2457066.7661690 | 20.2458 | 27.3322 | 20.8375 | 27.6592 | T24 | |
| 2457066.7205903 | 20.1083 | 27.2475 | 20.9042 | 27.6939 | T21 | |
| 2457067.6212963 | 20.3583 | 27.2225 | 20.9750 | 27.6189 | T21 | |
| 2457067.6606481 | 20.8417 | 27.5100 | 20.3292 | 27.2358 | T24 | |
| 2457068.2687384 | 19.9917 | 26.9606 | 20.3000 | 27.1297 | H | M13 |
| 2457068.2836921 | 20.2958 | 27.1233 | 20.6208 | 27.3044 | H | M13 |
| 2457068.3028009 | 20.6125 | 27.2939 | 20.9583 | 27.4811 | H | M13 |
| 2457068.3168287 | 20.9250 | 27.4608 | 21.2917 | 27.6597 | H | M13 |
| 2457068.3320370 | 21.2333 | 27.6256 | 21.5625 | 27.8081 | H | M13 |
| 2457069.2387731 | 19.6958 | 26.6567 | 20.1458 | 26.9064 | H | M14 |
| 2457069.2551273 | 20.0042 | 26.8267 | 20.4542 | 27.0736 | H | M14 |
| 2457069.2717014 | 20.3083 | 26.990 | 20.7583 | 27.2347 | H | M14 |
| 2457069.2882292 | 20.6208 | 27.1617 | 21.0708 | 27.4058 | H | M14 |
| 2457069.3040278 | 20.9958 | 27.3650 | 21.3750 | 27.5647 | H | M14 |
| 2457069.3196875 | 21.2417 | 27.4911 | 21.6833 | 27.7283 | H | M14 |
| 2457069.3359491 | 21.5542 | 27.6544 | 22.0000 | 27.8947 | H | M14 |
| 2457091.2913426 | 25.9875 | 27.2353 | 26.4167 | 27.4444 | H | M15 |
| 2457093.5936806 | 26.1375 | 27.0811 | 26.8083 | 27.3892 | T11 | |
| 2457253.0797685 | 72.5292 | 40.4531 | 73.2250 | 40.4589 | T24 | M10 |
| 2457256.9498380 | 72.9583 | 40.9542 | 73.6500 | 40.9528 | T24 | |
| 2457257.4351852 | 73.0542 | 41.0078 | 73.5583 | 41.0147 | H | M16 |
| 2457319.0221644 | 57.6875 | 49.8689 | 58.5000 | 49.9108 | T24 | |
| 2457319.9596875 | 57.6792 | 49.9394 | 58.4958 | 49.9933 | T24 | M11 |

**Summary of Abbreviations (used facilities)**

T30 is located at the Siding Spring Observatory, Australia. OTA: Planewave 20" Corrected Dall-Kirkham Astrograph with focal length of 2249 mm. F/Ratio is f/6.8. CCD: FLI-PL6303E Non-Anti Blooming Gate. Filter: AstroDon Tru-Balance Gen 2 E series Luminance. Guiding: Planewave Ascension HR200.

T24 is located at the Sierra Remote Observatory, Auberry, California, USA. OTA: Planewave 24" Corrected Dall-Kirkham Astrograph with focal length of 3962 mm. F/Ratio is f/6.5. CCD: FLI-PL09000 Anti Blooming Gate. Filter AstroDon E series Luminance. Guiding: Planewave Ascension HR200.

T11 is located at the New Mexico Skies Observatory at Mayhill, New Mexico USA. OTA: Planewave 20" Corrected Dall-Kirkham Astrograph with focal length of 2280 with focal reducer. F/Ratio is f/4.5. CCD: FLI ProLine PL11002M Anti Blooming Gate. Filter AstroDon Luminance. Guiding: Planewave Ascension HR200.

T21 is located at the New Mexico Skies Observatory at Mayhill, New Mexico, USA. OTA: Planewave 17" Corrected Dall-Kirkham Astrograph with focal length of 1940mm with focal reducer. F/Ratio f/4.5. CCD: FLI-PL6303E Non-Anti Blooming Gate. Filter Luminance. Guiding: Planewave Ascension 200HR.

H is located at the Hankasalmi Observatory, Finland. Telescope in the Hankasalmi Observatory is 40 cm RC Optical Systems 16RC with Ritchey-Chretien Cassegrain optics and Paramount ME mount. CCD camera is SBIG STX-16803 with photometric filters. The observatory is under remote control from internet.

**Calculating the equatorial coordinates of the convergence point**

The calculation of the celestial position of the convergence point of the dust trail is based on the simulation performed for 2021 September 6. The narrowest dust trail point was estimated from the plot of the calculated celestial coordinates and the heliocentric Cartesian coordinates of that point was used in the calculation.

The Earth's heliocentric Cartesian coordinates for the time interval 2021 January 1 to 2022 December 31 at 00 UT was produced using JPL/NASA Horizons Solar System Dynamics system web interface (Giorgin, 2021) at a one day interval.

The final coordinate calculations were performed with the same PHP script that had successfully produced the 2013, 2014 and 2015 coordinates (Lyytinen et al., 2013; 2015). The convergence point Cartesian coordinates were transformed to Geocentric coordinates for each day and were then converted to the equatorial coordinates. The maximum convergence point coordinate error is estimated to be ±3' in declination.

**Table S2:** Predicted coordinates of the dust trail nearest section to the 2007 explosion point for years 2021 and 2022.

| RA (hms) | DEC (dms) | Date 00 UT |
|---|---|---|
| 01h 06m 26s | 32° 45' 04" | 2021-Jan-01 00 UT |
| 01h 06m 10s | 32° 30' 58" | |
| 01h 05m 57s | 32° 17' 13" | |
| 01h 05m 46s | 32° 03' 51" | |
| 01h 05m 38s | 31° 50' 50" | |
| 01h 05m 32s | 31° 38' 11" | |
| 01h 05m 28s | 31° 25' 53" | |
| 01h 05m 26s | 31° 13' 56" | |
| 01h 05m 27s | 31° 02' 20" | |
| 01h 05m 30s | 30° 51' 05" | |
| 01h 05m 34s | 30° 40' 11" | 2021-Jan-11 00 UT |
| 01h 05m 41s | 30° 29' 37" | |
| 01h 05m 50s | 30° 19' 23" | |
| 01h 06m 01s | 30° 09' 30" | |
| 01h 06m 13s | 29° 59' 56" | |
| 01h 06m 27s | 29° 50' 41" | |
| 01h 06m 43s | 29° 41' 46" | |
| 01h 07m 01s | 29° 33' 10" | |
| 01h 07m 20s | 29° 24' 53" | |
| 01h 07m 41s | 29° 16' 54" | |
| 01h 08m 03s | 29° 09' 13" | 2021-Jan-21 00 UT |
| 01h 08m 27s | 29° 01' 49" | |
| 01h 08m 53s | 28° 54' 43" | |
| 01h 09m 19s | 28° 47' 55" | |
| 01h 09m 47s | 28° 41' 23" | |
| 01h 10m 17s | 28° 35' 07" | |
| 01h 10m 47s | 28° 29' 08" | |
| 01h 11m 19s | 28° 23' 24" | |
| 01h 11m 53s | 28° 17' 57" | |
| 01h 12m 27s | 28° 12' 44" | |
| 01h 13m 02s | 28° 07' 46" | 2021-Jan-31 00 UT |
| 01h 13m 39s | 28° 03' 03" | |
| 01h 14m 17s | 27° 58' 35" | |
| 01h 14m 56s | 27° 54' 20" | |
| 01h 15m 35s | 27° 50' 19" | |
| 01h 16m 16s | 27° 46' 32" | |
| 01h 16m 58s | 27° 42' 59" | |
| 01h 17m 41s | 27° 39' 38" | |
| 01h 18m 25s | 27° 36' 30" | |
| 01h 19m 09s | 27° 33' 35" | |
| 01h 19m 55s | 27° 30' 52" | 2021-Feb-10 00 UT |
| 01h 20m 42s | 27° 28' 21" | |
| 01h 21m 29s | 27° 26' 02" | |
| 01h 22m 17s | 27° 23' 54" | |
| 01h 23m 06s | 27° 21' 58" | |
| 01h 23m 56s | 27° 20' 13" | |
| 01h 24m 46s | 27° 18' 39" | |
| 01h 25m 37s | 27° 17' 15" | |
| 01h 26m 29s | 27° 16' 02" | |
| 01h 27m 22s | 27° 14' 59" | |
| 01h 28m 15s | 27° 14' 06" | 2021-Feb-20 00 UT |
| 01h 29m 09s | 27° 13' 22" | |
| 01h 30m 04s | 27° 12' 48" | |
| 01h 30m 59s | 27° 12' 23" | |
| 01h 31m 55s | 27° 12' 07" | |
| 01h 32m 52s | 27° 12' 00" | |
| 01h 33m 49s | 27° 12' 02" | |
| 01h 34m 46s | 27° 12' 12" | |
| 01h 35m 45s | 27° 12' 30" | |
| 01h 36m 43s | 27° 12' 56" | |
| 01h 37m 43s | 27° 13' 30" | 2021-Mar-02 00 UT |
| 01h 38m 43s | 27° 14' 11" | |
| 01h 39m 43s | 27° 15' 00" | |
| 01h 40m 44s | 27° 15' 56" | |
| 01h 41m 45s | 27° 17' 00" | |
| 01h 42m 47s | 27° 18' 10" | |
| 01h 43m 49s | 27° 19' 27" | |
| 01h 44m 52s | 27° 20' 51" | |
| 01h 45m 55s | 27° 22' 22" | |
| 01h 46m 59s | 27° 23' 59" | |
| 01h 48m 03s | 27° 25' 42" | 2021-Mar-12 00 UT |
| 01h 49m 07s | 27° 27' 31" | |
| 01h 50m 12s | 27° 29' 26" | |
| 01h 51m 17s | 27° 31' 27" | |
| 01h 52m 23s | 27° 33' 34" | |
| 01h 53m 29s | 27° 35' 46" | |
| 01h 54m 35s | 27° 38' 04" | |
| 01h 55m 42s | 27° 40' 27" | |
| 01h 56m 49s | 27° 42' 55" | |
| 01h 57m 57s | 27° 45' 28" | |
| 01h 59m 04s | 27° 48' 06" | 2021-Mar-22 00 UT |
| 02h 00m 12s | 27° 50' 48" | |
| 02h 01m 21s | 27° 53' 35" | |
| 02h 02m 30s | 27° 56' 27" | |
| 02h 03m 39s | 27° 59' 22" | |
| 02h 04m 48s | 28° 02' 22" | |
| 02h 05m 57s | 28° 05' 26" | |
| 02h 07m 07s | 28° 08' 35" | |
| 02h 08m 17s | 28° 11' 46" | |
| 02h 09m 28s | 28° 15' 02" | |
| 02h 10m 38s | 28° 18' 22" | 2021-Apr-01 00 UT |
| 02h 11m 49s | 28° 21' 45" | |
| 02h 13m 00s | 28° 25' 11" | |
| 02h 14m 12s | 28° 28' 41" | |
| 02h 15m 23s | 28° 32' 15" | |
| 02h 16m 35s | 28° 35' 51" | |
| 02h 17m 47s | 28° 39' 31" | |
| 02h 18m 59s | 28° 43' 14" | |
| 02h 20m 12s | 28° 47' 00" | |
| 02h 21m 25s | 28° 50' 49" | |
| 02h 22m 38s | 28° 54' 41" | 2021-Apr-11 00 UT |
| 02h 23m 51s | 28° 58' 36" | |
| 02h 25m 04s | 29° 02' 33" | |
| 02h 26m 17s | 29° 06' 33" | |
| 02h 27m 31s | 29° 10' 36" | |
| 02h 28m 45s | 29° 14' 41" | |
| 02h 29m 59s | 29° 18' 48" | |
| 02h 31m 13s | 29° 22' 58" | |
| 02h 32m 27s | 29° 27' 10" | |
| 02h 33m 42s | 29° 31' 24" | |
| 02h 34m 56s | 29° 35' 40" | 2021-Apr-21 00 UT |
| 02h 36m 11s | 29° 39' 58" | |
| 02h 37m 26s | 29° 44' 18" | |
| 02h 38m 41s | 29° 48' 40" | |
| 02h 39m 56s | 29° 53' 04" | |
| 02h 41m 11s | 29° 57' 29" | |
| 02h 42m 27s | 30° 01' 57" | |
| 02h 43m 42s | 30° 06' 25" | |
| 02h 44m 58s | 30° 10' 56" | |
| 02h 46m 13s | 30° 15' 28" | |
| 02h 47m 29s | 30° 20' 02" | 2021-May-01 00 UT |
| 02h 48m 45s | 30° 24' 37" | |
| 02h 50m 01s | 30° 29' 13" | |
| 02h 51m 17s | 30° 33' 51" | |
| 02h 52m 33s | 30° 38' 31" | |
| 02h 53m 49s | 30° 43' 12" | |
| 02h 55m 05s | 30° 47' 54" | |
| 02h 56m 21s | 30° 52' 37" | |
| 02h 57m 38s | 30° 57' 22" | |
| 02h 58m 54s | 31° 02' 08" | |
| 03h 00m 11s | 31° 06' 54" | 2021-May-11 00 UT |
| 03h 01m 27s | 31° 11' 43" | |
| 03h 02m 44s | 31° 16' 32" | |
| 03h 04m 00s | 31° 21' 22" | |
| 03h 05m 17s | 31° 26' 13" | |
| 03h 06m 33s | 31° 31' 05" | |
| 03h 07m 50s | 31° 35' 58" | |
| 03h 09m 06s | 31° 40' 52" | |
| 03h 10m 23s | 31° 45' 47" | |
| 03h 11m 40s | 31° 50' 42" | |
| 03h 12m 56s | 31° 55' 39" | 2021-May-21 00 UT |
| 03h 14m 13s | 32° 00' 36" | |
| 03h 15m 29s | 32° 05' 34" | |
| 03h 16m 46s | 32° 10' 32" | |
| 03h 18m 02s | 32° 15' 31" | |
| 03h 19m 19s | 32° 20' 31" | |
| 03h 20m 35s | 32° 25' 32" | |
| 03h 21m 51s | 32° 30' 33" | |
| 03h 23m 08s | 32° 35' 35" | |
| 03h 24m 24s | 32° 40' 38" | |
| 03h 25m 40s | 32° 45' 42" | 2021-May-31 00 UT |
| 03h 26m 56s | 32° 50' 46" | |
| 03h 28m 12s | 32° 55' 50" | |
| 03h 29m 28s | 33° 00' 56" | |
| 03h 30m 44s | 33° 06' 02" | |
| 03h 32m 00s | 33° 11' 09" | |
| 03h 33m 16s | 33° 16' 16" | |
| 03h 34m 32s | 33° 21' 24" | |
| 03h 35m 47s | 33° 26' 33" | |
| 03h 37m 02s | 33° 31' 42" | |
| 03h 38m 18s | 33° 36' 52" | 2021-Jun-10 00 UT |
| 03h 39m 33s | 33° 42' 03" | |
| 03h 40m 48s | 33° 47' 14" | |
| 03h 42m 03s | 33° 52' 26" | |
| 03h 43m 17s | 33° 57' 38" | |
| 03h 44m 32s | 34° 02' 51" | |
| 03h 45m 46s | 34° 08' 05" | |
| 03h 47m 00s | 34° 13' 19" | |
| 03h 48m 14s | 34° 18' 34" | |
| 03h 49m 28s | 34° 23' 50" | |
| 03h 50m 41s | 34° 29' 07" | 2021-Jun-20 00 UT |
| 03h 51m 55s | 34° 34' 24" | |
| 03h 53m 08s | 34° 39' 41" | |
| 03h 54m 21s | 34° 45' 00" | |
| 03h 55m 33s | 34° 50' 19" | |
| 03h 56m 46s | 34° 55' 39" | |
| 03h 57m 58s | 35° 01' 00" | |
| 03h 59m 10s | 35° 06' 22" | |
| 04h 00m 21s | 35° 11' 44" | |
| 04h 01m 33s | 35° 17' 08" | |
| 04h 02m 44s | 35° 22' 33" | 2021-Jun-30 00 UT |
| 04h 03m 54s | 35° 27' 58" | |
| 04h 05m 05s | 35° 33' 25" | |
| 04h 06m 15s | 35° 38' 52" | |
| 04h 07m 24s | 35° 44' 21" | |
| 04h 08m 34s | 35° 49' 51" | |
| 04h 09m 43s | 35° 55' 22" | |
| 04h 10m 52s | 36° 00' 54" | |
| 04h 12m 00s | 36° 06' 28" | |
| 04h 13m 08s | 36° 12' 02" | |
| 04h 14m 15s | 36° 17' 39" | 2021-Jul-10 00 UT |
| 04h 15m 22s | 36° 23' 16" | |
| 04h 16m 29s | 36° 28' 55" | |
| 04h 17m 35s | 36° 34' 35" | |
| 04h 18m 41s | 36° 40' 17" | |
| 04h 19m 46s | 36° 46' 01" | |
| 04h 20m 51s | 36° 51' 46" | |
| 04h 21m 55s | 36° 57' 32" | |
| 04h 22m 58s | 37° 03' 20" | |
| 04h 24m 01s | 37° 09' 10" | |
| 04h 25m 04s | 37° 15' 02" | 2021-Jul-20 00 UT |
| 04h 26m 06s | 37° 20' 56" | |
| 04h 27m 07s | 37° 26' 52" | |
| 04h 28m 08s | 37° 32' 49" | |
| 04h 29m 08s | 37° 38' 49" | |
| 04h 30m 08s | 37° 44' 51" | |
| 04h 31m 07s | 37° 50' 55" | |
| 04h 32m 05s | 37° 57' 02" | |
| 04h 33m 02s | 38° 03' 11" | |
| 04h 33m 59s | 38° 09' 22" | |
| 04h 34m 55s | 38° 15' 36" | 2021-Jul-30 00 UT |
| 04h 35m 50s | 38° 21' 53" | |
| 04h 36m 45s | 38° 28' 12" | |
| 04h 37m 38s | 38° 34' 34" | |
| 04h 38m 31s | 38° 40' 59" | |
| 04h 39m 23s | 38° 47' 27" | |
| 04h 40m 14s | 38° 53' 58" | |
| 04h 41m 05s | 39° 00' 32" | |
| 04h 41m 54s | 39° 07' 10" | |
| 04h 42m 42s | 39° 13' 50" | |
| 04h 43m 29s | 39° 20' 34" | 2021-Aug-09 00 UT |
| 04h 44m 16s | 39° 27' 21" | |
| 04h 45m 01s | 39° 34' 12" | |
| 04h 45m 45s | 39° 41' 07" | |
| 04h 46m 28s | 39° 48' 05" | |
| 04h 47m 10s | 39° 55' 06" | |
| 04h 47m 50s | 40° 02' 12" | |
| 04h 48m 30s | 40° 09' 22" | |
| 04h 49m 08s | 40° 16' 35" | |
| 04h 49m 45s | 40° 23' 53" | |
| 04h 50m 20s | 40° 31' 14" | 2021-Aug-19 00 UT |
| 04h 50m 55s | 40° 38' 40" | |
| 04h 51m 28s | 40° 46' 10" | |
| 04h 51m 59s | 40° 53' 45" | |
| 04h 52m 29s | 41° 01' 24" | |
| 04h 52m 57s | 41° 09' 08" | |
| 04h 53m 24s | 41° 16' 56" | |
| 04h 53m 49s | 41° 24' 50" | |
| 04h 54m 13s | 41° 32' 48" | |
| 04h 54m 35s | 41° 40' 50" | |
| 04h 54m 55s | 41° 48' 58" | 2021-Aug-29 00 UT |
| 04h 55m 13s | 41° 57' 11" | |
| 04h 55m 30s | 42° 05' 28" | |
| 04h 55m 44s | 42° 13' 51" | |
| 04h 55m 57s | 42° 22' 19" | |
| 04h 56m 07s | 42° 30' 52" | |
| 04h 56m 16s | 42° 39' 29" | |
| 04h 56m 22s | 42° 48' 12" | |
| 04h 56m 26s | 42° 57' 00" | |
| 04h 56m 28s | 43° 05' 53" | |
| 04h 56m 27s | 43° 14' 51" | 2021-Sep-08 00 UT |
| 04h 56m 24s | 43° 23' 53" | |
| 04h 56m 18s | 43° 33' 01" | |
| 04h 56m 10s | 43° 42' 13" | |
| 04h 55m 59s | 43° 51' 30" | |
| 04h 55m 46s | 44° 00' 51" | |
| 04h 55m 29s | 44° 10' 16" | |
| 04h 55m 10s | 44° 19' 45" | |
| 04h 54m 48s | 44° 29' 19" | |
| 04h 54m 22s | 44° 38' 56" | |
| 04h 53m 54s | 44° 48' 37" | 2021-Sep-18 00 UT |
| 04h 53m 22s | 44° 58' 21" | |
| 04h 52m 48s | 45° 08' 07" | |
| 04h 52m 09s | 45° 17' 57" | |
| 04h 51m 28s | 45° 27' 49" | |
| 04h 50m 42s | 45° 37' 42" | |
| 04h 49m 53s | 45° 47' 38" | |
| 04h 49m 00s | 45° 57' 34" | |
| 04h 48m 04s | 46° 07' 30" | |
| 04h 47m 03s | 46° 17' 27" | |
| 04h 45m 59s | 46° 27' 22" | 2021-Sep-28 00 UT |
| 04h 44m 50s | 46° 37' 17" | |
| 04h 43m 37s | 46° 47' 09" | |
| 04h 42m 20s | 46° 56' 58" | |
| 04h 40m 58s | 47° 06' 43" | |
| 04h 39m 32s | 47° 16' 24" | |
| 04h 38m 01s | 47° 25' 59" | |
| 04h 36m 25s | 47° 35' 27" | |
| 04h 34m 45s | 47° 44' 47" | |
| 04h 33m 00s | 47° 53' 59" | |
| 04h 31m 09s | 48° 03' 00" | 2021-Oct-08 00 UT |
| 04h 29m 14s | 48° 11' 50" | |
| 04h 27m 14s | 48° 20' 27" | |
| 04h 25m 09s | 48° 28' 50" | |
| 04h 22m 59s | 48° 36' 58" | |
| 04h 20m 44s | 48° 44' 48" | |
| 04h 18m 23s | 48° 52' 20" | |
| 04h 15m 58s | 48° 59' 32" | |
| 04h 13m 27s | 49° 06' 21" | |
| 04h 10m 52s | 49° 12' 48" | |
| 04h 08m 11s | 49° 18' 49" | 2021-Oct-18 00 UT |
| 04h 05m 26s | 49° 24' 24" | |
| 04h 02m 35s | 49° 29' 30" | |
| 03h 59m 40s | 49° 34' 06" | |
| 03h 56m 40s | 49° 38' 10" | |
| 03h 53m 36s | 49° 41' 39" | |
| 03h 50m 27s | 49° 44' 34" | |
| 03h 47m 14s | 49° 46' 51" | |
| 03h 43m 57s | 49° 48' 29" | |
| 03h 40m 37s | 49° 49' 26" | |
| 03h 37m 12s | 49° 49' 41" | 2021-Oct-28 00 UT |
| 03h 33m 45s | 49° 49' 13" | |
| 03h 30m 14s | 49° 48' 00" | |
| 03h 26m 41s | 49° 46' 00" | |
| 03h 23m 05s | 49° 43' 14" | |
| 03h 19m 28s | 49° 39' 39" | |
| 03h 15m 48s | 49° 35' 15" | |
| 03h 12m 07s | 49° 30' 01" | |
| 03h 08m 25s | 49° 23' 57" | |
| 03h 04m 43s | 49° 17' 03" | |
| 03h 01m 00s | 49° 09' 18" | 2021-Nov-07 00 UT |
| 02h 57m 17s | 49° 00' 43" | |
| 02h 53m 35s | 48° 51' 18" | |
| 02h 49m 54s | 48° 41' 03" | |
| 02h 46m 13s | 48° 30' 00" | |
| 02h 42m 35s | 48° 18' 09" | |
| 02h 38m 58s | 48° 05' 32" | |
| 02h 35m 24s | 47° 52' 09" | |
| 02h 31m 52s | 47° 38' 02" | |
| 02h 28m 23s | 47° 23' 12" | |
| 02h 24m 57s | 47° 07' 42" | 2021-Nov-17 00 UT |
| 02h 21m 35s | 46° 51' 34" | |
| 02h 18m 16s | 46° 34' 48" | |
| 02h 15m 01s | 46° 17' 28" | |
| 02h 11m 50s | 45° 59' 35" | |
| 02h 08m 43s | 45° 41' 11" | |
| 02h 05m 41s | 45° 22' 19" | |
| 02h 02m 43s | 45° 03' 02" | |
| 01h 59m 50s | 44° 43' 20" | |
| 01h 57m 01s | 44° 23' 18" | |
| 01h 54m 17s | 44° 02' 57" | 2021-Nov-27 00 UT |
| 01h 51m 39s | 43° 42' 19" | |
| 01h 49m 05s | 43° 21' 27" | |
| 01h 46m 36s | 43° 00' 23" | |
| 01h 44m 12s | 42° 39' 10" | |
| 01h 41m 53s | 42° 17' 49" | |
| 01h 39m 40s | 41° 56' 23" | |
| 01h 37m 31s | 41° 34' 54" | |
| 01h 35m 27s | 41° 13' 24" | |
| 01h 33m 28s | 40° 51' 54" | |
| 01h 31m 34s | 40° 30' 28" | 2021-Dec-07 00 UT |
| 01h 29m 45s | 40° 09' 06" | |
| 01h 28m 00s | 39° 47' 50" | |
| 01h 26m 20s | 39° 26' 42" | |
| 01h 24m 45s | 39° 05' 43" | |
| 01h 23m 15s | 38° 44' 56" | |
| 01h 21m 48s | 38° 24' 20" | |
| 01h 20m 26s | 38° 03' 57" | |
| 01h 19m 09s | 37° 43' 48" | |
| 01h 17m 55s | 37° 23' 54" | |
| 01h 16m 46s | 37° 04' 17" | 2021-Dec-17 00 UT |
| 01h 15m 40s | 36° 44' 57" | |
| 01h 14m 39s | 36° 25' 54" | |
| 01h 13m 41s | 36° 07' 10" | |
| 01h 12m 47s | 35° 48' 44" | |
| 01h 11m 57s | 35° 30' 38" | |
| 01h 11m 10s | 35° 12' 52" | |
| 01h 10m 26s | 34° 55' 26" | |
| 01h 09m 46s | 34° 38' 21" | |
| 01h 09m 09s | 34° 21' 37" | |
| 01h 08m 35s | 34° 05' 14" | 2021-Dec-27 00 UT |
| 01h 08m 04s | 33° 49' 13" | |
| 01h 07m 36s | 33° 33' 33" | |
| 01h 07m 11s | 33° 18' 15" | |
| 01h 06m 49s | 33° 03' 19" | |
| 01h 06m 30s | 32° 48' 45" | |
| 01h 06m 14s | 32° 34' 32" | |
| 01h 06m 00s | 32° 20' 42" | |
| 01h 05m 48s | 32° 07' 14" | |
| 01h 05m 39s | 31° 54' 07" | |
| 01h 05m 32s | 31° 41' 22" | 2022-Jan-06 00 UT |
| 01h 05m 28s | 31° 28' 59" | |
| 01h 05m 26s | 31° 16' 57" | |
| 01h 05m 26s | 31° 05' 16" | |
| 01h 05m 29s | 30° 53' 57" | |
| 01h 05m 33s | 30° 42' 58" | |
| 01h 05m 39s | 30° 32' 19" | |
| 01h 05m 48s | 30° 22' 01" | |
| 01h 05m 58s | 30° 12' 03" | |
| 01h 06m 10s | 30° 02' 25" | |
| 01h 06m 24s | 29° 53' 06" | 2022-Jan-16 00 UT |
| 01h 06m 39s | 29° 44' 06" | |
| 01h 06m 56s | 29° 35' 26" | |
| 01h 07m 15s | 29° 27' 03" | |
| 01h 07m 35s | 29° 18' 59" | |
| 01h 07m 57s | 29° 11' 14" | |
| 01h 08m 21s | 29° 03' 45" | |
| 01h 08m 46s | 28° 56' 34" | |
| 01h 09m 12s | 28° 49' 41" | |
| 01h 09m 40s | 28° 43' 04" | |
| 01h 10m 09s | 28° 36' 43" | 2022-Jan-26 00 UT |
| 01h 10m 39s | 28° 30' 39" | |
| 01h 11m 11s | 28° 24' 51" | |
| 01h 11m 43s | 28° 19' 19" | |
| 01h 12m 17s | 28° 14' 02" | |
| 01h 12m 53s | 28° 9' 00" | |
| 01h 13m 29s | 28° 04' 13" | |
| 01h 14m 07s | 27° 59' 41" | |
| 01h 14m 45s | 27° 55' 23" | |
| 01h 15m 25s | 27° 51' 19" | |
| 01h 16m 06s | 27° 47' 29" | 2022-Feb-05 00 UT |
| 01h 16m 47s | 27° 43' 52" | |
| 01h 17m 30s | 27° 40' 29" | |
| 01h 18m 13s | 27° 37' 18" | |
| 01h 18m 58s | 27° 34' 20" | |
| 01h 19m 43s | 27° 31' 34" | |
| 01h 20m 30s | 27° 29' 01" | |
| 01h 21m 17s | 27° 26' 39" | |
| 01h 22m 05s | 27° 24' 29" | |
| 01h 22m 53s | 27° 22' 30" | |
| 01h 23m 43s | 27° 20' 42" | 2022-Feb-15 00 UT |
| 01h 24m 33s | 27° 19' 05" | |
| 01h 25m 24s | 27° 17' 39" | |
| 01h 26m 16s | 27° 16' 23" | |
| 01h 27m 08s | 27° 15' 17" | |
| 01h 28m 01s | 27° 14' 20" | |
| 01h 28m 55s | 27° 13' 34" | |
| 01h 29m 50s | 27° 12' 57" | |
| 01h 30m 45s | 27° 12' 29" | |
| 01h 31m 40s | 27° 12' 11" | |
| 01h 32m 37s | 27° 12' 01" | 2022-Feb-25 00 UT |
| 01h 33m 34s | 27° 12' 00" | |
| 01h 34m 31s | 27° 12' 07" | |
| 01h 35m 29s | 27° 12' 23" | |
| 01h 36m 28s | 27° 12' 47" | |
| 01h 37m 27s | 27° 13' 19" | |
| 01h 38m 27s | 27° 13' 58" | |
| 01h 39m 27s | 27° 14' 46" | |
| 01h 40m 28s | 27° 15' 41" | |
| 01h 41m 29s | 27° 16' 42" | |
| 01h 42m 31s | 27° 17' 52" | 2022-Mar-07 00 UT |
| 01h 43m 33s | 27° 19' 08" | |
| 01h 44m 36s | 27° 20' 30" | |
| 01h 45m 39s | 27° 22' 00" | |
| 01h 46m 42s | 27° 23' 35" | |
| 01h 47m 46s | 27° 25' 17" | |
| 01h 48m 51s | 27° 27' 05" | |
| 01h 49m 55s | 27° 28' 59" | |
| 01h 51m 01s | 27° 30' 58" | |
| 01h 52m 06s | 27° 33' 04" | |
| 01h 53m 12s | 27° 35' 14" | 2022-Mar-17 00 UT |
| 01h 54m 18s | 27° 37' 30" | |
| 01h 55m 25s | 27° 39' 52" | |
| 01h 56m 32s | 27° 42' 18" | |
| 01h 57m 39s | 27° 44' 49" | |
| 01h 58m 47s | 27° 47' 25" | |
| 01h 59m 55s | 27° 50' 06" | |
| 02h 01m 03s | 27° 52' 52" | |
| 02h 02m 12s | 27° 55' 42" | |
| 02h 03m 21s | 27° 58' 36" | |
| 02h 04m 30s | 28° 01' 35" | 2022-Mar-27 00 UT |
| 02h 05m 39s | 28° 04' 38" | |
| 02h 06m 49s | 28° 07' 45" | |
| 02h 07m 59s | 28° 10' 56" | |
| 02h 09m 09s | 28° 14' 11" | |
| 02h 10m 20s | 28° 17' 30" | |
| 02h 11m 31s | 28° 20' 52" | |
| 02h 12m 42s | 28° 24' 18" | |
| 02h 13m 53s | 28° 27' 48" | |
| 02h 15m 05s | 28° 31' 21" | |
| 02h 16m 17s | 28° 34' 57" | 2022-Apr-06 00 UT |
| 02h 17m 29s | 28° 38' 36" | |
| 02h 18m 41s | 28° 42' 19" | |
| 02h 19m 54s | 28° 46' 05" | |
| 02h 21m 06s | 28° 49' 53" | |
| 02h 22m 19s | 28° 53' 44" | |
| 02h 23m 32s | 28° 57' 38" | |
| 02h 24m 45s | 29° 01' 35" | |
| 02h 25m 59s | 29° 05' 34" | |
| 02h 27m 12s | 29° 09' 36" | |
| 02h 28m 26s | 29° 13' 40" | 2022-Apr-16 00 UT |
| 02h 29m 40s | 29° 17' 46" | |
| 02h 30m 54s | 29° 21' 55" | |
| 02h 32m 08s | 29° 26' 06" | |
| 02h 33m 23s | 29° 30' 19" | |
| 02h 34m 37s | 29° 34' 34" | |
| 02h 35m 52s | 29° 38' 52" | |
| 02h 37m 07s | 29° 43' 11" | |
| 02h 38m 22s | 29° 47' 32" | |
| 02h 39m 37s | 29° 51' 56" | |
| 02h 40m 52s | 29° 56' 21" | 2022-Apr-26 00 UT |
| 02h 42m 07s | 30° 00' 48" | |
| 02h 43m 23s | 30° 05' 16" | |
| 02h 44m 38s | 30° 09' 47" | |
| 02h 45m 54s | 30° 14' 19" | |
| 02h 47m 10s | 30° 18' 52" | |
| 02h 48m 26s | 30° 23' 27" | |
| 02h 49m 42s | 30° 28' 04" | |
| 02h 50m 58s | 30° 32' 42" | |
| 02h 52m 14s | 30° 37' 21" | |
| 02h 53m 30s | 30° 42' 02" | 2022-May-06 00 UT |
| 02h 54m 46s | 30° 46' 44" | |
| 02h 56m 02s | 30° 51' 27" | |
| 02h 57m 19s | 30° 56' 12" | |
| 02h 58m 35s | 31° 00' 57" | |
| 02h 59m 51s | 31° 05' 44" | |
| 03h 01m 08s | 31° 10' 31" | |
| 03h 02m 24s | 31° 15' 20" | |
| 03h 03m 41s | 31° 20' 10" | |
| 03h 04m 57s | 31° 25' 00" | |
| 03h 06m 14s | 31° 29' 52" | 2022-May-16 00 UT |
| 03h 07m 30s | 31° 34' 44" | |
| 03h 08m 47s | 31° 39' 38" | |
| 03h 10m 03s | 31° 44' 32" | |
| 03h 11m 20s | 31° 49' 27" | |
| 03h 12m 37s | 31° 54' 23" | |
| 03h 13m 53s | 31° 59' 20" | |
| 03h 15m 10s | 32° 04' 17" | |
| 03h 16m 26s | 32° 09' 16" | |
| 03h 17m 43s | 32° 14' 15" | |
| 03h 18m 59s | 32° 19' 15" | 2022-May-26 00 UT |
| 03h 20m 16s | 32° 24' 16" | |
| 03h 21m 32s | 32° 29' 17" | |
| 03h 22m 49s | 32° 34' 19" | |
| 03h 24m 05s | 32° 39' 22" | |
| 03h 25m 21s | 32° 44' 26" | |
| 03h 26m 37s | 32° 49' 30" | |
| 03h 27m 54s | 32° 54' 35" | |
| 03h 29m 10s | 32° 59' 40" | |
| 03h 30m 26s | 33° 04' 46" | |
| 03h 31m 41s | 33° 09' 53" | 2022-Jun-05 00 UT |
| 03h 32m 57s | 33° 15' 00" | |
| 03h 34m 13s | 33° 20' 08" | |
| 03h 35m 28s | 33° 25' 17" | |
| 03h 36m 44s | 33° 30' 26" | |
| 03h 37m 59s | 33° 35' 35" | |
| 03h 39m 14s | 33° 40' 46" | |
| 03h 40m 29s | 33° 45' 57" | |
| 03h 41m 44s | 33° 51' 08" | |
| 03h 42m 58s | 33° 56' 20" | |
| 03h 44m 13s | 34° 01' 33" | 2022-Jun-15 00 UT |
| 03h 45m 27s | 34° 06' 46" | |
| 03h 46m 41s | 34° 12' 00" | |
| 03h 47m 55s | 34° 17' 15" | |
| 03h 49m 09s | 34° 22' 30" | |
| 03h 50m 23s | 34° 27' 46" | |
| 03h 51m 36s | 34° 33' 03" | |
| 03h 52m 49s | 34° 38' 21" | |
| 03h 54m 02s | 34° 43' 39" | |
| 03h 55m 15s | 34° 48' 58" | |
| 03h 56m 28s | 34° 54' 18" | 2022-Jun-25 00 UT |
| 03h 57m 40s | 34° 59' 39" | |
| 03h 58m 52s | 35° 05' 01" | |
| 04h 00m 04s | 35° 10' 24" | |
| 04h 01m 15s | 35° 15' 48" | |
| 04h 02m 26s | 35° 21' 12" | |
| 04h 03m 37s | 35° 26' 38" | |
| 04h 04m 47s | 35° 32' 04" | |
| 04h 05m 58s | 35° 37' 32" | |
| 04h 07m 07s | 35° 43' 00" | |
| 04h 08m 17s | 35° 48' 30" | 2022-Jul-05 00 UT |
| 04h 09m 26s | 35° 54' 00" | |
| 04h 10m 35s | 35° 59' 32" | |
| 04h 11m 43s | 36° 05' 05" | |
| 04h 12m 51s | 36° 10' 40" | |
| 04h 13m 58s | 36° 16' 15" | |
| 04h 15m 05s | 36° 21' 52" | |
| 04h 16m 12s | 36° 27' 30" | |
| 04h 17m 18s | 36° 33' 10" | |
| 04h 18m 24s | 36° 38' 51" | |
| 04h 19m 29s | 36° 44' 34" | 2022-Jul-15 00 UT |
| 04h 20m 34s | 36° 50' 18" | |
| 04h 21m 38s | 36° 56' 04" | |
| 04h 22m 42s | 37° 01' 52" | |
| 04h 23m 46s | 37° 07' 41" | |
| 04h 24m 48s | 37° 13' 33" | |
| 04h 25m 50s | 37° 19' 26" | |
| 04h 26m 52s | 37° 25' 21" | |
| 04h 27m 53s | 37° 31' 19" | |
| 04h 28m 53s | 37° 37' 18" | |
| 04h 29m 53s | 37° 43' 20" | 2022-Jul-25 00 UT |
| 04h 30m 52s | 37° 49' 24" | |
| 04h 31m 51s | 37° 55' 30" | |
| 04h 32m 48s | 38° 01' 39" | |
| 04h 33m 45s | 38° 07' 50" | |
| 04h 34m 41s | 38° 14' 03" | |
| 04h 35m 37s | 38° 20' 19" | |
| 04h 36m 32s | 38° 26' 38" | |
| 04h 37m 25s | 38° 33' 00" | |
| 04h 38m 18s | 38° 39' 24" | |
| 04h 39m 11s | 38° 45' 51" | 2022-Aug-04 00 UT |
| 04h 40m 02s | 38° 52' 21" | |
| 04h 40m 52s | 38° 58' 54" | |
| 04h 41m 42s | 39° 05' 31" | |
| 04h 42m 30s | 39° 12' 10" | |
| 04h 43m 17s | 39° 18' 53" | |
| 04h 44m 04s | 39° 25' 39" | |
| 04h 44m 49s | 39° 32' 28" | |
| 04h 45m 34s | 39° 39' 22" | |
| 04h 46m 17s | 39° 46' 19" | |
| 04h 46m 59s | 39° 53' 19" | 2022-Aug-14 00 UT |
| 04h 47m 40s | 40° 00' 24" | |
| 04h 48m 20s | 40° 07' 32" | |
| 04h 48m 59s | 40° 14' 44" | |
| 04h 49m 36s | 40° 22' 01" | |
| 04h 50m 12s | 40° 29' 22" | |
| 04h 50m 46s | 40° 36' 47" | |
| 04h 51m 20s | 40° 44' 16" | |
| 04h 51m 52s | 40° 51' 50" | |
| 04h 52m 22s | 40° 59' 28" | |
| 04h 52m 51s | 41° 07' 11" | 2022-Aug-24 00 UT |
| 04h 53m 18s | 41° 14' 59" | |
| 04h 53m 44s | 41° 22' 51" | |
| 04h 54m 08s | 41° 30' 48" | |
| 04h 54m 30s | 41° 38' 50" | |
| 04h 54m 51s | 41° 46' 57" | |
| 04h 55m 09s | 41° 55' 08" | |
| 04h 55m 26s | 42° 03' 25" | |
| 04h 55m 41s | 42° 11' 46" | |
| 04h 55m 54s | 42° 20' 12" | |
| 04h 56m 05s | 42° 28' 44" | 2022-Sep-03 00 UT |
| 04h 56m 14s | 42° 37' 20" | |
| 04h 56m 21s | 42° 46' 01" | |
| 04h 56m 25s | 42° 54' 47" | |
| 04h 56m 27s | 43° 03' 39" | |
| 04h 56m 27s | 43° 12' 35" | |
| 04h 56m 25s | 43° 21' 36" | |
| 04h 56m 20s | 43° 30' 41" | |
| 04h 56m 12s | 43° 39' 52" | |
| 04h 56m 02s | 43° 49' 07" | |
| 04h 55m 49s | 43° 58' 27" | 2022-Sep-13 00 UT |
| 04h 55m 34s | 44° 07' 51" | |
| 04h 55m 15s | 44° 17' 20" | |
| 04h 54m 54s | 44° 26' 53" | |
| 04h 54m 30s | 44° 36' 29" | |
| 04h 54m 02s | 44° 46' 09" | |
| 04h 53m 31s | 44° 55' 53" | |
| 04h 52m 57s | 45° 05' 39" | |
| 04h 52m 20s | 45° 15' 29" | |
| 04h 51m 39s | 45° 25' 21" | |
| 04h 50m 55s | 45° 35' 14" | 2022-Sep-23 00 UT |
| 04h 50m 07s | 45° 45' 10" | |
| 04h 49m 15s | 45° 55' 06" | |
| 04h 48m 19s | 46° 05' 02" | |
| 04h 47m 20s | 46° 14' 59" | |
| 04h 46m 16s | 46° 24' 55" | |
| 04h 45m 08s | 46° 34' 50" | |
| 04h 43m 56s | 46° 44' 42" | |
| 04h 42m 40s | 46° 54' 32" | |
| 04h 41m 19s | 47° 04' 17" | |
| 04h 39m 54s | 47° 13' 59" | 2022-Oct-03 00 UT |
| 04h 38m 24s | 47° 23' 35" | |
| 04h 36m 49s | 47° 33' 04" | |
| 04h 35m 10s | 47° 42' 26" | |
| 04h 33m 26s | 47° 51' 39" | |
| 04h 31m 38s | 48° 00' 43" | |
| 04h 29m 44s | 48° 09' 35" | |
| 04h 27m 45s | 48° 18' 16" | |
| 04h 25m 41s | 48° 26' 42" | |
| 04h 23m 33s | 48° 34' 54" | |
| 04h 21m 19s | 48° 42' 49" | 2022-Oct-13 00 UT |
| 04h 19m 00s | 48° 50' 26" | |
| 04h 16m 36s | 48° 57' 44" | |
| 04h 14m 07s | 49° 04' 41" | |
| 04h 11m 33s | 49° 11' 14" | |
| 04h 08m 53s | 49° 17' 23" | |
| 04h 06m 09s | 49° 23' 05" | |
| 04h 03m 20s | 49° 28' 19" | |
| 04h 00m 26s | 49° 33' 04" | |
| 03h 57m 27s | 49° 37' 16" | |
| 03h 54m 24s | 49° 40' 55" | 2022-Oct-23 00 UT |
| 03h 51m 16s | 49° 43' 58" | |
| 03h 48m 04s | 49° 46' 25" | |
| 03h 44m 48s | 49° 48' 13" | |
| 03h 41m 28s | 49° 49' 20" | |
| 03h 38m 04s | 49° 49' 45" | |
| 03h 34m 38s | 49° 49' 27" | |
| 03h 31m 08s | 49° 48' 25" | |
| 03h 27m 35s | 49° 46' 36" | |
| 03h 24m 00s | 49° 44' 01" | |
| 03h 20m 23s | 49° 40' 37" | 2022-Nov-02 00 UT |
| 03h 16m 44s | 49° 36' 25" | |
| 03h 13m 03s | 49° 31' 24" | |
| 03h 09m 22s | 49° 25' 32" | |
| 03h 05m 39s | 49° 18' 51" | |
| 03h 01m 57s | 49° 11' 20" | |
| 02h 58m 14s | 49° 02' 58" | |
| 02h 54m 32s | 48° 53' 47" | |
| 02h 50m 50s | 48° 43' 46" | |
| 02h 47m 10s | 48° 32' 56" | |
| 02h 43m 31s | 48° 21' 18" | 2022-Nov-12 00 UT |
| 02h 39m 54s | 48° 08' 54" | |
| 02h 36m 19s | 47° 55' 43" | |
| 02h 32m 46s | 47° 41' 48" | |
| 02h 29m 17s | 47° 27' 10" | |
| 02h 25m 50s | 47° 11' 50" | |
| 02h 22m 26s | 46° 55' 52" | |
| 02h 19m 06s | 46° 39' 15" | |
| 02h 15m 50s | 46° 22' 03" | |
| 02h 12m 38s | 46° 04' 17" | |
| 02h 09m 30s | 45° 46' 00" | 2022-Nov-22 00 UT |
| 02h 06m 26s | 45° 27' 14" | |
| 02h 03m 27s | 45° 08' 02" | |
| 02h 00m 33s | 44° 48' 25" | |
| 01h 57m 43s | 44° 28' 27" | |
| 01h 54m 58s | 44° 08' 09" | |
| 01h 52m 18s | 43° 47' 35" | |
| 01h 49m 43s | 43° 26' 45" | |
| 01h 47m 13s | 43° 05' 44" | |
| 01h 44m 48s | 42° 44' 33" | |
| 01h 42m 28s | 42° 23' 14" | 2022-Dec-02 00 UT |
| 01h 40m 13s | 42° 01' 50" | |
| 01h 38m 03s | 41° 40' 22" | |
| 01h 35m 58s | 41° 18' 53" | |
| 01h 33m 58s | 40° 57' 24" | |
| 01h 32m 03s | 40° 35' 58" | |
| 01h 30m 13s | 40° 14' 36" | |
| 01h 28m 27s | 39° 53' 20" | |
| 01h 26m 46s | 39° 32' 10" | |
| 01h 25m 10s | 39° 11' 10" | |
| 01h 23m 38s | 38° 50' 19" | 2022-Dec-12 00 UT |
| 01h 22m 10s | 38° 29' 40" | |
| 01h 20m 47s | 38° 09' 14" | |
| 01h 19m 28s | 37° 49' 01" | |
| 01h 18m 14s | 37° 29' 03" | |
| 01h 17m 03s | 37° 09' 21" | |
| 01h 15m 57s | 36° 49' 56" | |
| 01h 14m 54s | 36° 30' 47" | |
| 01h 13m 55s | 36° 11' 57" | |
| 01h 13m 00s | 35° 53' 26" | |
| 01h 12m 09s | 35° 35' 14" | 2022-Dec-22 00 UT |
| 01h 11m 21s | 35° 17' 22" | |
| 01h 10m 36s | 34° 59' 50" | |
| 01h 09m 55s | 34° 42' 39" | |
| 01h 09m 17s | 34° 25' 49" | |
| 01h 08m 43s | 34° 09' 21" | |
| 01h 08m 11s | 33° 53' 14" | |
| 01h 07m 43s | 33° 37' 29" | |
| 01h 07m 17s | 33° 22' 07" | |
| 01h 06m 55s | 33° 07' 06" | |

**Table S3:** The phase angle values (in degrees) calculated for the trail narrowest point for each JD in the model. The values are calculated in this work with the described here new Dust Trail kit model.

| JD | abbreviation | phase angle |
|---|---|---|
| 2457067 | M1 | 21.77288 |
| 2456341 | M2 | 15.55373 |
| 2456529 | M3 | 13.64544 |
| 2456700 | M4 | 12.32833 |
| 2456897 | M5 | 23.56685 |
| 2456917 | M6 | 24.12191 |
| 2457018 | M7 | 21.33922 |
| 2457043 | M8 | 23.52844 |
| 2457044 | M9 | 23.98090 |
| 2457253 | M10 | 22.44840 |
| 2457320 | M11 | 16.64098 |
| 2456907 | M12 | 24.64619 |
| 2457068 | M13 | 21.58406 |
| 2457069 | M14 | 21.45343 |
| 2457091 | M15 | 17.50541 |
| 2457257 | M16 | 22.64547 |
| 2459454 | F1 | 22.35707 |
| 2459464 | F2 | 22.55161 |
| 2459474 | F3 | 22.44405 |
| 2459600 | F4 | 22.59722 |
| 2459601 | F5 | 22.58704 |
| 2459624 | F6 | 21.51010 |
| 2459626 | F7 | 21.25446 |
| 2459648 | F8 | 18.12913 |
| 2459810 | F9 | 20.84181 |

**Image subtraction routine**

Stack both night's images separately and save them to the TIFF file. Then stack each with a fixed field, which has an artificial register file. Save to TIFF. Change each of them to negative and stack two resulting files.

**Figures**

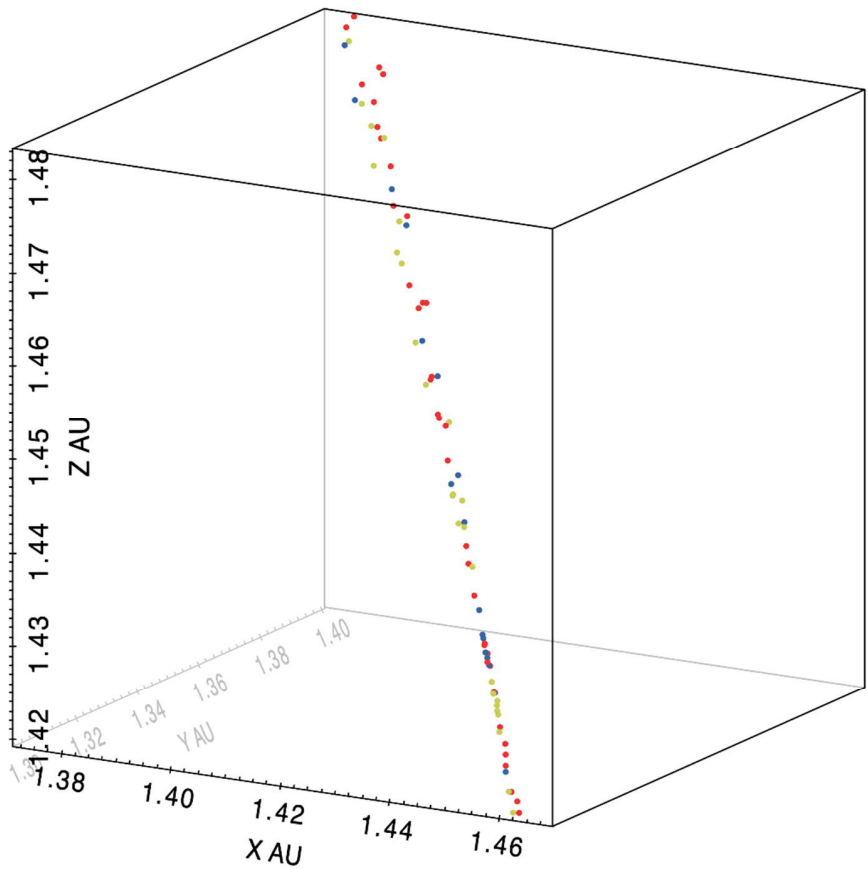

Figure S1. Modeled dust trail in February 2015 (2015-02-14T12:00:00) at the nearest point of the 2007 outburst. The particles are shown in the ICRF coordinates XYZ. Colour code of modeled particles: Blue: SPs. Yellow: MPs. Red: BPs.

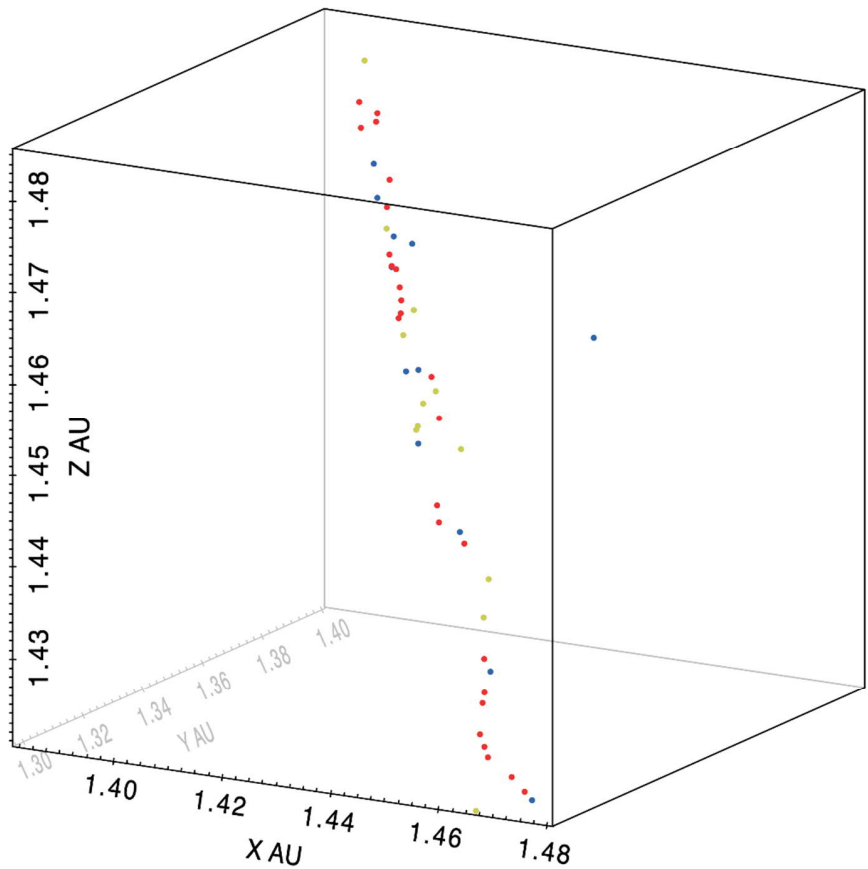

Figure S2. Modeled dust trail in February 2022 (2022-02-15T12:00:00) at the nearest point of the 2007 outburst point. The particles are shown in the ICRF coordinates XYZ. Colour code of modeled particles: Blue: SPs. Yellow: MPs. Red: BPs.

Figure S3. Modeling of the dust trail in August 2021 (2021-08-27T12:00:00) (F1). The X-axis shows RA and the Y-axis DEC. Blue: SPs. Yellow: MPs. Red: BPs. Particles ejected towards the Sun are marked with crosses.

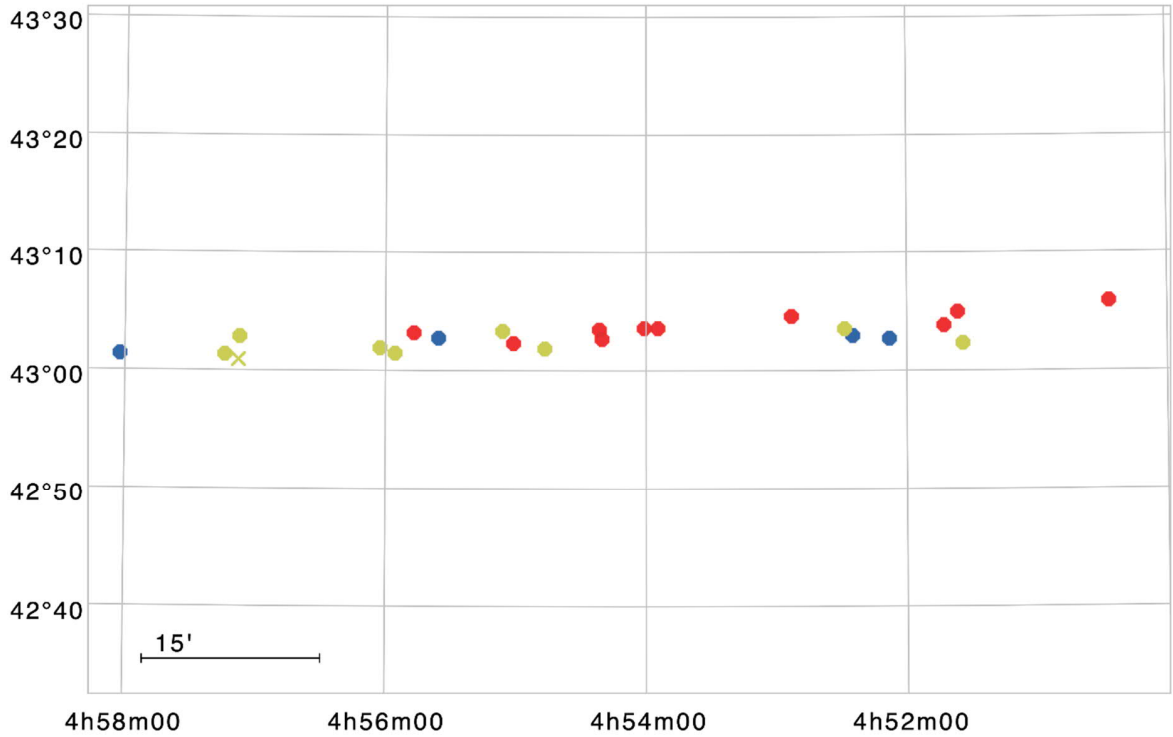

Figure S4. Modeling of the dust trail in September 2021 (2021-09-06T12:00:00) (F2). The X-axis shows RA and the Y-axis DEC. Blue: SPs. Yellow: MPs. Red: BPs. Particles ejected towards the Sun are marked with crosses.

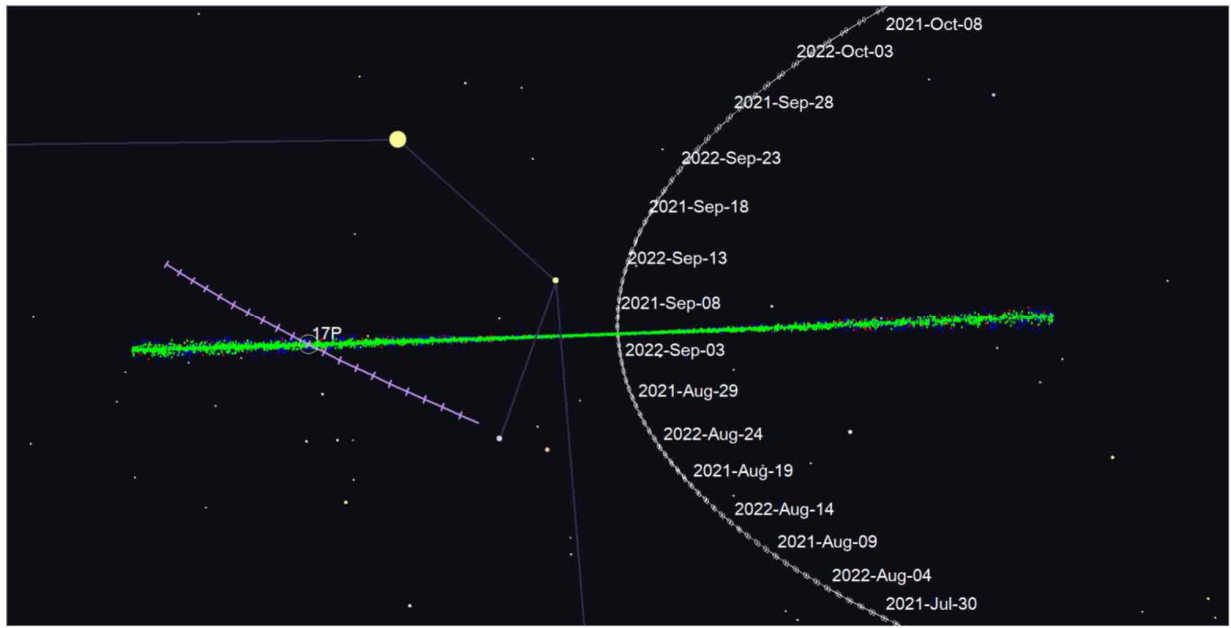

Figure S5. Comet 17P/Holmes plotted on top of the modeled trail for 2021 September 6. The convergence point location movement in the sky is also shown.

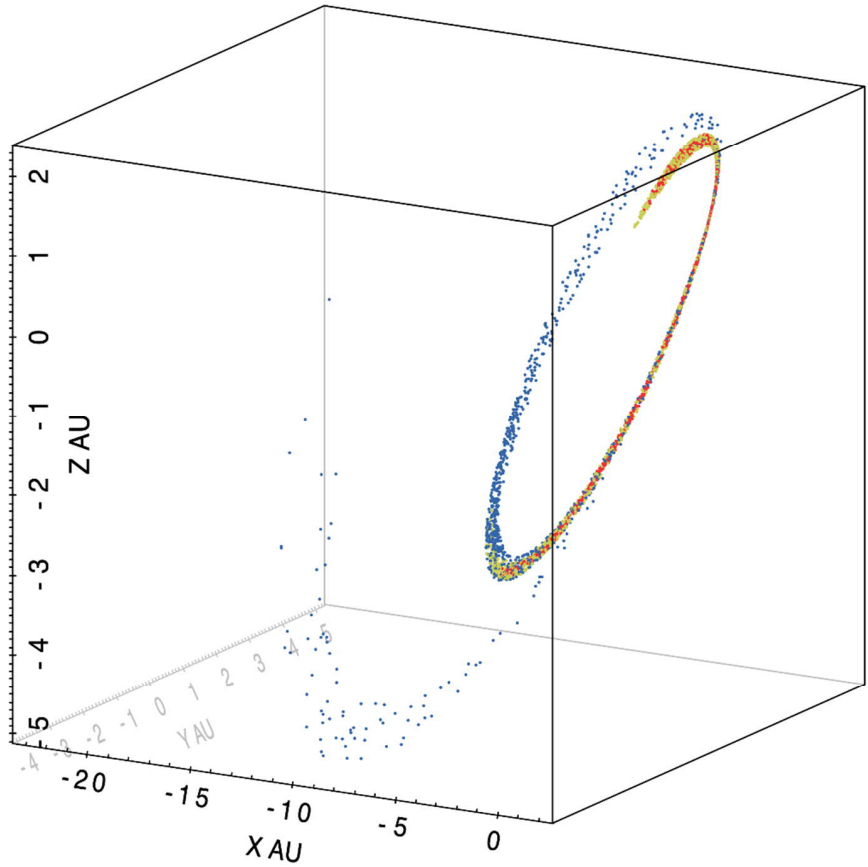

Figure S6. Dust trail prediction modeled for 2021 October 28. The outburst point convergence hourglass pattern is at the forward section of the trail. The southern sky node convergence hourglass pattern is at the tail section. Particles are shown in the ICRF coordinates XYZ. Blue: SPs. Yellow: MPs. Red: BPs.

**DEDICATION**

We dedicate this paper to the memory of mastermind Esko Lyytinen and his life-long passion for science. From a very early age Esko was fascinated by the stars and the universe. Esko was 14 when the Earth's first artificial satellite, Sputnik 1, was launched in 1957 and already at that age he could figure out exactly where and when to look to see it in the sky. Esko continued observing Sputnik 1 as well as subsequent satellites. It may come as no surprise then that years later he named his first model of meteoroid stream formation "the satellite model of comets" (Lyytinen E., 1999).

Esko had exceptional talent for mathematics and was very keen on its applications in orbital mechanics. He did a tremendous amount of modeling and calculations for meteor stream predictions for the scientific community. The most important stream for him to model was Leonids. Then, when 17P/Holmes underwent the outburst, he immediately noticed the potential for dust trail visibility even in modest telescopes and he envisioned that the brightness would be such to make the dust trail visible even without a telescope. While this did not happen, the opportunity to directly observe this kind of phenomenon is fascinating.

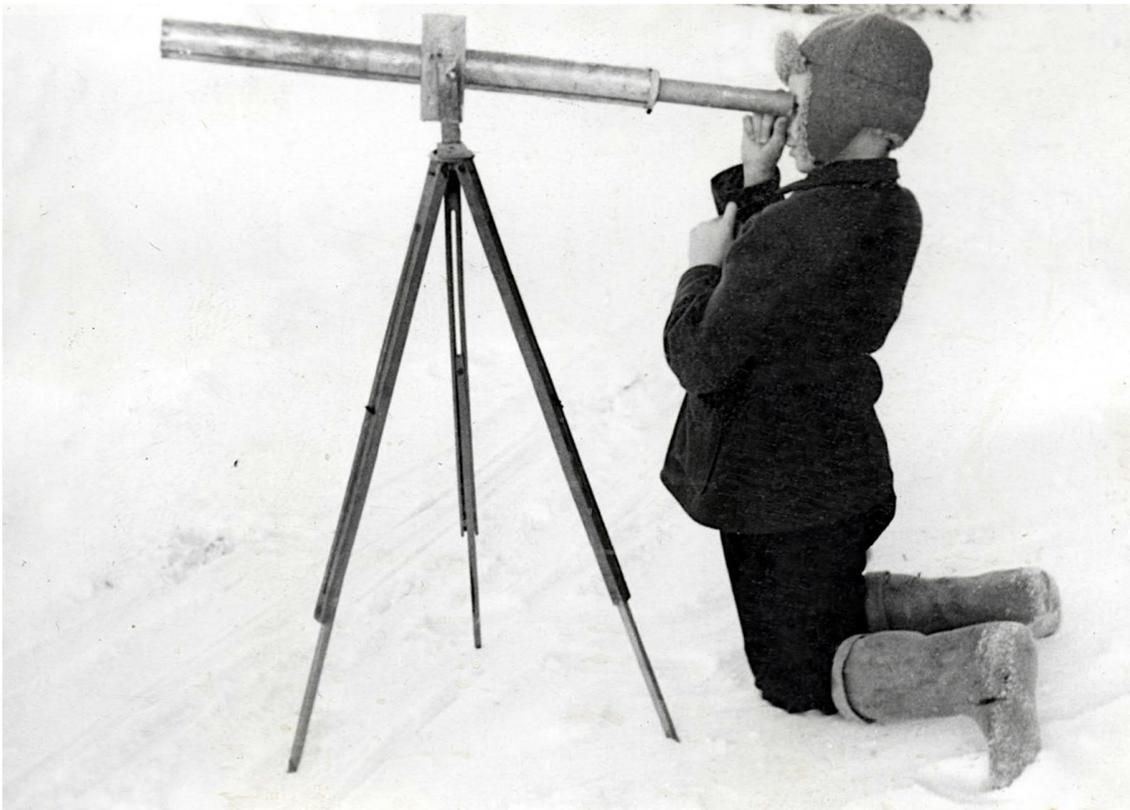

Esko on the photograph titled "my first telescope" (source: Esko's own photograph album)